\documentclass[10pt,conference]{IEEEtran}
\IEEEoverridecommandlockouts
\usepackage{algorithm,,algpseudocode,tabularx} 
\makeatletter
\newcommand{\multiline}[1]{%
  \begin{tabularx}{\dimexpr\linewidth-\ALG@thistlm}[t]{@{}X@{}}
    #1
  \end{tabularx}
}
\makeatother
\usepackage{cite}
\usepackage[utf8x]{inputenc} 
\usepackage{amsmath}
\usepackage{bm}
\usepackage{amssymb,amsmath}
\usepackage[mathscr]{euscript}
\usepackage{graphicx}
\graphicspath{ {./Figures/} }
\usepackage{epsfig}
\usepackage{grffile}
\usepackage{balance,color}
\usepackage{xcolor, soul}
\usepackage{framed}
\usepackage{lipsum}
\colorlet{shadecolor}{yellow!20}

\providecommand{\keywords}[1]{{\textit{Index Terms}}}
\abovecaptionskip 
\belowcaptionskip
\usepackage{caption}
\usepackage{subcaption} 
\usepackage{multirow}

\begin{document}

\title{Slicing for AI: An Online Learning Framework for Network Slicing Supporting AI Services }  

\author{Menna Helmy$^1$, Alaa Awad Abdellatif$^1$, Naram Mhaisen$^2$, Amr Mohamed$^1$, Aiman Erbad$^1$,  \\
\begin{tabular}{c}
$^1$ College of Engineering, Qatar University, Qatar \\ 
$^2$ College of Electrical Engineering, Mathematics, and Computer Science, TU Delft,  Netherlands \\ 
Email: \{mh1800882,  amrm\}@qu.edu.qa, \{alaa.abdellatif, aerbad\}@ieee.org\\ 
		\thanks { This work was made possible by NPRP grant \#  NPRP13S-0205-200265 from the Qatar National Research Fund (a member of Qatar Foundation). The findings achieved herein are solely the responsibility of the authors.  }  
\end{tabular}
}
\maketitle

\begin{abstract}  
   The forthcoming 6G networks will embrace a new realm of AI-driven services that requires innovative network slicing strategies, namely \textit{slicing for AI}, which involves the creation of customized network slices to meet Quality of service (QoS) requirements of diverse AI services. This poses challenges due to time-varying dynamics of users' behavior and mobile networks. Thus, this paper proposes an online learning framework to optimize the allocation of computational and communication resources to AI services, while considering their unique key performance indicators (KPIs), such as accuracy, latency, and cost. 
   We define a problem of optimizing the total accuracy while balancing conflicting KPIs, prove its NP-hardness, and propose an online learning framework for solving it in dynamic environments.  
   We present a basic online solution and two variations employing a pre-learning elimination method for reducing the decision space to expedite the learning. Furthermore, we propose a biased decision space subset selection by incorporating prior knowledge to enhance the learning speed without compromising performance and present two alternatives of handling the selected subset. Our results depict the efficiency of the proposed solutions in converging to the optimal decisions, while reducing decision space and improving time complexity.   
   \end{abstract}


\section{Introduction\label{sec:Introduction}} 

It is anticipated that the 6G networks will have the capability to cater to an array of services, each with distinct quality of service (QoS) specifications. Such services may include multisensory extended reality, autonomous driving, and hologram video streaming. To ensure diversity in services, similar to the 5G networks, network slicing is employed to create multiple slices for various services over a shared physical network infrastructure. An economical management strategy for network slicing can enable the fulfillment of QoS requirements throughout the different phases of the lifecycle, such as preparation, planning, and scheduling \cite{wu2022ai}. Additionally, newly emerging technologies for 6G networks such as the Open Radio Access Network (O-RAN) architecture will allow the integration of native Artificial intelligence (AI) solutions to accommodate heterogeneous service deployments \cite{ORAN,azariah22survey,polese22survey}.
In this context, AI will become omnipresent in 6G networks, meaning that it will penetrate every aspect of the network, creating a state of ubiquitous intelligence. Network nodes will possess in-built AI capabilities, not only allowing for intelligent network management but also promoting the growth of AI-based services, such as machine learning, natural language processing, and computer vision.

To fulfil diverse requirements of different AI services, it is important to implement customized network slices, namely \textit{slicing for AI}. This approach allows for the creation of tailored network slices that can cater to the distinct requirements of various AI services (i.e., accuracy, learning speed, etc.), thereby optimizing the network's overall performance\cite{AI21}. Indeed, slicing for AI refers to the creation of  network slices with customized resources to meet QoS requirements of diverse AI services. 
Hence, it can play a crucial role in the process of dynamically distributing and assigning resources in meta learning systems by  allocating resources such as computation power, memory, and data samples efficiently and effectively to improve the learning process and overall performance of the meta learning algorithms.  
However, this problem is challenging due to the dynamics of diverse learning algorithms and mobile networks. For example, the quality and distribution of acquired data are time-varying and heavily affect the performance of AI services. Hence, novel solutions are needed to adapt to such non-stationary dynamics.   
Thus,  this paper aims to optimize the decision-making process by proposing online learning techniques \cite{OL12} that continuously observe the system's performance without prior knowledge of the expected behavior.  


Several works have addressed the problem of network slicing in 5G and beyond networks to support the heterogeneous requirements of various conventional vertical services, such as ultra-reliable low-latency communication (URLLC), enhanced mobile boradband (eMBB), and massive machine-type communication (mMTC) \cite{DynamicRAN,FDRL,ORANslicing,VERA,DynSlicingConf}. However,  
few works considered the problem of \textit{slicing for AI}. Most of the proposed solutions in this context leveraged offline classical optimization and reinforcement learning (RL) techniques. 
Classical optimization approaches assume full knowledge about the behavior of the environment while leveraging mathematical formulations to model different objectives and constraints. However, such an assumption may be impractical with some services and mobile network systems, which are highly dynamic and time-variant. On the contrary, RL approaches aim to learn a policy in a stateful system by mapping between states and actions. These approaches consider an environment that is stationary (i.e., behaving according to a non-varying distribution). Also, classical optimization and the training phase of RL are usually performed in an offline manner on an available dataset/environment. Nonetheless, an offline approach may fail in adapting to dynamic/time-varying systems. 
Thus, in this paper, we propose an online learning framework to address the problem of \textit{slicing for AI}. The proposed  framework can adapt to different system dynamics and uncertainty with regret guarantees. The main contributions of this paper are as follows: 
\begin{itemize}
\item We formulate the problem of \textit{slicing for AI} services as an optimization problem with the objective of maximizing performance, which turns out to be NP-hard. 

\item We introduce an online learning framework and propose a basic online solution to solve the formulated problem. We propose two alternatives of the solution where each adopts a pre-learning decision space reduction process to expedite the learning. The first alternative merges similar decisions to obtain a compact structure of the decision space, while the second builds upon the obtained compact structure and identifies candidates of the optimal decision to optimize the size of the action space.
\item We propose a subset selection of the original decision space with prior knowledge to accelerate the convergence of the solution. We consider two alternative approaches of manipulating the solution by biasing the selected subset.


\item We assess the solution's performance by comparing it to optimal allocation and fixed allocation benchmarks, demonstrating its ability to converge towards optimal resource allocation. We examine the complexity and convergence of the proposed alternative online solutions. Furthermore, we compare between the trade-offs of the two biased subset selection approaches.

\end{itemize} 
 
The rest of the paper is organized as follows. We present some related work in Section \ref{sec:Related}. 
Section \ref{sec:System_and_problem} introduces the system model and the formulation of the problem as an optimization problem.
Section \ref{sec:Solution} presents the proposed solutions and modelling the problem into an online learning framework. Section \ref{sec:Simulation} evaluates the performance of the solution. Finally, we conclude the paper in Section \ref{sec:conclusion}. 

\section{Related work \label{sec:Related}}
In this section, we review existing literature that offers diverse solutions for network slicing in 5G and beyond networks, including optimization, RL, and online learning solutions.   

\textbf{Optimization-based solutions:} Several studies have tackled the issue of network slicing and resource allocation using traditional optimization techniques \cite{ORANslicing,Zeng20FLrrm,edgeCloudCont,DML22,Lin21AI}. For instance, the authors in \cite{ORANslicing} addressed the problem of allocating base-band resources for diverse slices in an O-RAN system, while considering the diverse requirements of eMBB, URLLC, and mMTC services. They formulated this problem as a mixed-integer non-linear programming problem and proposed a service-aware solution through a two-step iterative heuristic algorithm. Similarly, in \cite{DML22}, the authors modeled their resource allocation problem as a mixed-integer non-linear programming problem and suggested approximation-based methods. The authors of \cite{Lin21AI} introduced an Integer Linear Programming algorithm to solve their formulated slicing problem. Optimization-based solutions rely on precise mathematical models and, although capable of achieving optimal or near-optimal solutions, they often face practical limitations due to the dynamic nature of mobile networks. Consequently, any changes in the network environment necessitate re-executing the optimization process to obtain new optimal solutions. In some cases, even the underlying mathematical model requires recalibration, resulting in computational overhead and hinders the practicality of the solution.

\textbf{RL-based solutions:} To overcome the limitations of classical optimization solutions, emerging studies have employed Deep RL (DRL) techniques for network slicing  \cite{DynamicRAN, abdellatif2023intelligent,  FDRL,HDRL21,VERA,DRL22,DRL21,ORAN-DRL22,DRL20,Guan21,ORAN-DRL23}. For example, the authors in \cite{VERA} proposed a multi-agent RL framework, namely VERA, for a fair Pareto-efficient allocation of computing and network resources to competing users' applications and virtual RAN (vRAN) services at the edge. Other works, such as \cite{FDRL, HDRL21}, addressed the problem of RAN slicing in 6G networks by proposing a Federated DRL (FDRL) and a DRL based solution respectively. The authors in \cite{DRL22,DRL21} proposed a DRL solution to the problem of resource allocation of eMBB and URLLC slices, subject to URLLC delay requirements and eMBB data rate requirements. Other network slicing problems such as RAN slice admission and placement, and slice reconfiguration has been addressed by \cite{ORAN-DRL22,DRL20} with a DRL-based solution. Although RL-based solutions have the advantage of adaptability to stationary system dynamics and the ability to maximize system rewards through policy learning, they often assume a non-varying stochastic environment, which can pose challenges when attempting to adapt to unforeseen changes in the system. 

\textbf{Online learning-based solutions:} In contrast to classical optimization and RL methods, online learning approaches offer the advantage of adaptability to non-stationary system dynamics and unforeseen events. However, only a limited number of studies have applied online learning in the context of network slicing  \cite{DynamicRAN,OCO}. For instance, the work in  \cite{DynamicRAN} incorporated online learning in solving its formulated Software-Defined Network (SDN) radio resource allocation problem. The objective of this work was to maximize the total achievable data rate of eMBB and URLLC end-users associated with all gNodeBs while adapting to user dynamics. The work in \cite{OCO} addressed the problem of resource reservation by service providers, which is closely related to network slicing problems. The authors proposed a framework based on online convex optimization for effective resource reservation while maximizing the services' performance, constrained to a time-average budget constraint, considering a time-varying service demand and slice pricing. 

\textbf{Slicing for AI:} While there have been several studies focusing on network slicing for conventional services, the number of works addressing the network slicing problem for AI-based services is comparatively limited \cite{Zeng20FLrrm,edgeCloudCont,DML22,Lin21AI}. For example, the authors in \cite{Zeng20FLrrm} tackled the challenge of joint bandwidth allocation and user scheduling for federated edge learning. Their objective was to reduce energy consumption while ensuring a guaranteed learning speed. Additionally, in \cite{edgeCloudCont}, the authors addressed the resource allocation problem for distributed machine learning (ML) application considering both edge and cloud resources in terms of computational performance and cost. In \cite{Lin21AI}, the authors focused on joint AI service placement and resource allocation in mobile edge system, while minimizing computation time and energy consumption. While these works contribute to resource allocation for AI services, none of them consider the joint allocation of resources and tuning of the underlying AI models' variables, such as the learning parameters (e.g., number of epochs) and training data size. %

\textbf{Novelty.} Our work is the first to tackle  the challenges of \textit{slicing for AI} through the utilization of online learning techniques. The proposed framework aims to optimize the performance of various AI services while effectively adapting to dynamic system conditions and unexpected events. Moreover, unlike other studies, we consider the joint allocation of computational and communication resources for different AI models along with hyper-parameter tuning of these models.     
\section{System model and problem formulation \label{sec:System_and_problem}}
In this section, we describe the considered system architecture, highlight the main functions constituting an AI service, discuss the key performance metrics that will be tackled for different AI services, and present the formulated Accuracy maximization problem. 

\subsection{System architecture\label{sec:Architecture}} 

We denote sets  with calligraphic capital letters, e.g., $\mathcal{A}$, and vectors with bold small letters , e.g., $\bm{a}$. We use the superscript $(t)$ to indicate the dependence on a time slot, e.g., $\bm{a}^{(t)}$, and subscripts to represent the correspondence to an indexed element. For example, to denote the $j$-th element of the set $\mathcal{A} =\{\bm{a}_{1}, \bm{a}_{2}, ..., \bm{a}_{J}\} $ we write $\bm{a}_{j}$, and to denote a parameter that corresponds to the $i$-th element of $\bm{a}_{j}$ we write $L_{i,j}$. To represent a sequence of vectors from slot $t = 1$ to slot $T$ we write $\{\bm{a}^{(t)}\}_{t=1}^{T}$. We present the key notation in Table \ref{tab:notation}.

{ \small
\renewcommand{\arraystretch}{1.2}
\begin{table}[t!]
	\centering
\caption{Key Notation}
	  \label{tab:notation}
\begin{tabular}{|c|c|} 
			\hline 
\textbf{Notation} & \textbf{Description} 
\\
\hline
$\mathcal{I}$ & The set of AI models

\\
\hline
$\mathcal{T}$ & The set of time slots 

\\
\hline
$L_{i}^{(t)}$ & The data size in samples required at the $t$-th slot
\\
&  for training AI model $i \in \mathcal{I}$
  \\
\hline
$l_{i}^{(t)}$ & The data size in percentage required at the $t$-th slot
\\
&  for training AI model $i \in \mathcal{I}$
  \\
\hline
$m_{i}^{(t)}$ & The number of epochs required at the $t$-th slot for
\\
 & training AI model $i \in \mathcal{I}$
\\
\hline
$\psi_{i}^{(t)}$ & The computing resources allocated at the $t$-th slot 
  \\
  & for training AI model $i \in \mathcal{I}$
\\
\hline
$\phi$ & The number of CPU cycles required to compute one 
\\
& sample of data
  \\
\hline
$\Psi_\text{max}$ & The maximum available computing resources
  \\
\hline
$\lambda_{i}^{(t)}$ &  The communication resources allocated at the $t$-th 
\\ 
& slot for training AI model $i \in \mathcal{I}$
\\
\hline
$\Lambda_\text{max}$ & The maximum available communication resources
  \\
\hline
$D_\text{max$_{i}$}$ & The maximum learning latency requested by the
\\
&  service corresponding to AI model $i \in \mathcal{I}$
  \\
\hline
$C_\text{max$_{i}$}$ & The maximum cost set by the service corresponding
\\
& to AI model $i \in \mathcal{I}$
  \\
\hline
$q_{i}^{(t)}(\cdot)$ & A function representing the learning accuracy  
\\
& of AI model $i \in \mathcal{I}$ at the $t$-th time slot
  \\
\hline
$\alpha_{i}$ & The priority of the service corresponding to
\\
&AI model $i \in \mathcal{I}$
  \\
\hline
$\bm{a}^{(t)}$ & Allocation decision vector selected at the $t$-th slot
  \\
\hline
$\mathcal{B}_\text{SA}^{(t)}$ & A super action defined as a set of grouped allocation 
\\
&decisions selected at the $t$-th slot 
  \\
\hline
\end{tabular} 
\end{table}
 }

\begin{figure*}[t!]
	\centering
		\scalebox{1.3}{\includegraphics[width=0.65 \textwidth, height = 6.7 cm ]{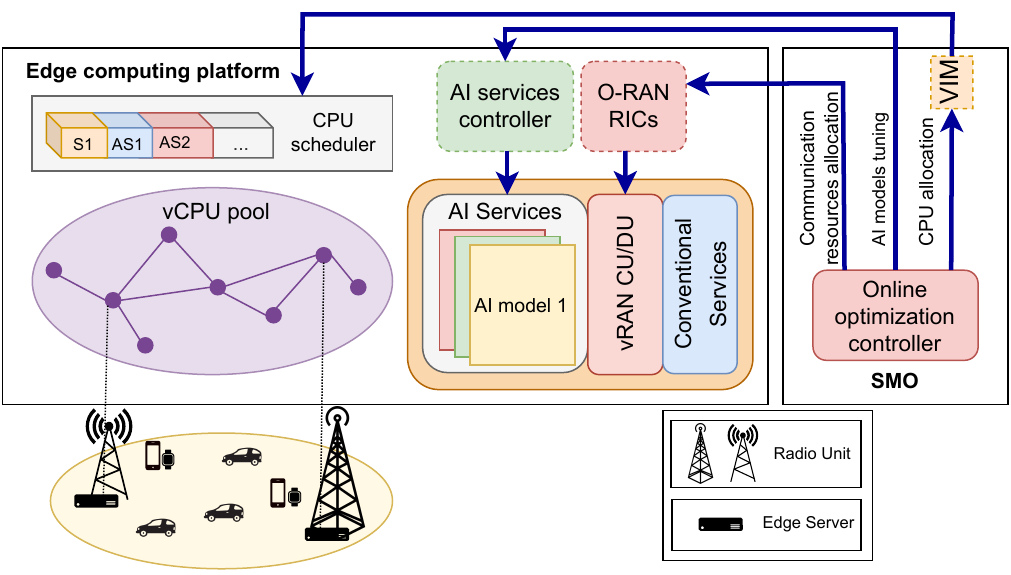}}
	\caption{System model under study.  
 }
	\label{fig:system_model}
\end{figure*}

The considered system model is depicted in
Figure \ref{fig:system_model}, which is adapted from \cite{VERA} and aligns with the O-RAN framework specifications \cite{ORAN}. The system supports conventional services, such as enhanced mobile broadband (eMBB) and ultra-reliable low-latency communication (URLLC), AI services, such as pandemic detection and object detection for autonomous driving services, and virtualized RAN (vRAN) functions, such as central unit (CU) and distributed unit (DU). Such services and functions coexist within an edge computing platform and compete over a shared pool of virtualized CPU (vCPU) resources (i.e., computing resources) managed by a CPU scheduler, as well as communication resources. 
We particularly consider the joint allocation of virtual resources to AI services represented by a set of underlying AI models, $\mathcal{I} = \{1,2,3,...,I\}$, and the tuning of some of the AI models parameters to perform AI model training. The system is time-slotted and consists of $T$ time slots, $t \in \mathcal{T} = \{1,2,...,T\}$. The online optimization controller running within the Service Management \& Orchestration (SMO) platform, represents a learner whose function is to control the allocation of resources and the tuning of the AI models parameters. The learner allocates computing and communication resources to the AI services through interaction with the virtual infrastructure manager (VIM), and the O-RAN intelligent controllers (RICs) respectively. It also tunes the AI models parameters through interacting with the AI services controllers.
Specifically, the learner selects a sequence of actions $\{\bm{a}^{(t)}\}_{t=1}^{T}$ corresponding to allocation decisions from a finite convex set of decisions/actions $\mathcal{A}$ at the beginning of each slot $t$, which results in a sequence of convex performance functions $\{f^{(t)}(\bm{a}^{(t)})\}_{t=1}^{T}$ that are only known at the end of $t$. 

\subsection{Key performance metrics \label{sec:Services}} 

 AI services, which comprises different functions supporting the continuous training and deployment of underlying AI models, exhibit their unique KPIs, including: learning speed, latency, and accuracy. 
The life cycle of an AI service can be divided into three primary phases or functions \cite{AI21}, which are:

\textbf{Data aggregation and data pre-processing:} 
    It involves collecting the user data necessary for training the learning model. The main resources in demand for this phase are communication resources. It is integral to jointly allocate the communication resources needed to serve this function and the computing resources needed to serve model training of each AI model, while fulfilling an allowable maximum latency requirement. 

 \textbf{Model training:} This phase is usually computationally intensive especially for complex AI models, and it depends on; the size of the training data, and the number of epochs. Tuning these two hyper-parameters for each AI model affects the computational demand as well as the performance of the trained model in terms of inference accuracy. 

\textbf{Model inference:} The performance of model inference is a crucial factor in obtaining the ideal allocation of resources and tuning of hyper-parameters for AI model training. If the AI model exhibits a low performance/accuracy, resource re-allocation and re-tuning of hyper-parameters may be necessary. By adjusting the resources allocated and tuning the hyper-parameters to perform AI model training functions, the AI model can be optimized to perform better while still maintaining the coexistence of multiple services/ AI models on the same platform.

Learning accuracy and latency are two essential performance metrics for any AI model $i \in \mathcal{I}$ that depend on; 1) the allocated computational and communication resources at the $t$-th time slot denoted by $\psi_{i}^{(t)}$ and $\lambda_{i}^{(t)}$ respectively, 2) the tuning of the two AI model training hyper-parameters, data size and number of epochs, at the $t$-th time slot which we denote by $L_{i}^{(t)}$ and $m_{i}^{(t)}$ respectively:

\textbf{Learning latency:} is defined as the total delay to fully train an AI model \cite{Pervasive22}, which consists of: the communication delay of acquiring data from different users/locations $d_\text{comm}$, and the processing delay for completing the model training $d_\text{proc}$. Therefore, the learning latency experienced by an AI model $i \in \mathcal{I}$ at the $t$-th slot is defined as:
\begin{equation}
    \label{eq:Time}
    \begin{split}
    D(L_{i}^{(t)}, m_{i}^{(t)}, \psi_{i}^{(t)}, \lambda_{i}^{(t)}) = d_\text{comm}(L_{i}^{(t)},\lambda_{i}^{(t)}) \\
    + d_\text{proc}(L_{i}^{(t)},m_{i}^{(t)},\psi_{i}^{(t)}), 
     \end{split}
\end{equation}
The communication delay at the $t$-th slot is defined as the time taken to transmit the acquired data of size $L_{i}^{(t)}$ from different locations to the designated server with the allocated data rate $\lambda_{i}^{(t)}$. It is expressed as:
\begin{equation}
    \label{eq:t_T}
    d_\text{comm}(L_{i}^{(t)},\lambda_{i}^{(t)}) = \frac{L_{i}^{(t)}}{\lambda_{i}^{(t)}} + \epsilon,
\end{equation}
where $\epsilon$ is the channel access delay \cite{awad2011energy}.  
The processing delay of training an AI model $i$ at the $t$-th slot is given by \cite{Pervasive22}:
\begin{equation}
    \label{eq:t_p}
    d_\text{proc}(L_{i}^{(t)},m_{i}^{(t)},\psi_{i}^{(t)}) = m_{i}^{(t)} \times (\frac{\phi L_{i}^{(t)}}{\psi_{i}^{(t)}}),
\end{equation}
where $m_{i}^{(t)}$ is the number of epochs, $\phi$ is the CPU cycles required to compute one sample of data, and $\psi_{i}^{(t)}$ is the CPU frequency allocated to service $i$ (i.e., computing resources). 
We remark that the allocated CPU, $\psi_{i}^{(t)}$, and data rate, $\lambda_{i}^{(t)}$, allow a degree of freedom with respect to the trade-off between the training latency and cost. In other words, allocating more resources would reduce the training latency, however, it would entail a higher cost. 

\textbf{Learning accuracy:} is defined as the inference (or prediction) result on input data compared against the true values, which directly reflects the quality of the training. We denote the inference accuracy of a model $i \in \mathcal{I}$ that has completed the training phase at slot $t$ as; $q_{i}^{(t)}(L_{i}^{(t)},m_{i}^{(t)}) \in [0,1]$. This accuracy depends on: 1) the size of the acquired data $L_{i}^{(t)}$, 2) the quality of the data, 3) the distribution of the training and inference data, 4) the number of epochs $m_{i}^{(t)}$ adopted to train the model, 5) and the service provider's decision regarding the method of training their AI model. Some methods of training AI models are explained in \cite{CT}: 1) \textit{full retraining}: involves periodic retraining of the model from scratch whenever a certain size of new data is available, 2) \textit{online training}: the model is initially trained and a mini-batch iteration is triggered once a new batch is available, 3) \textit{proactive training}: a training iteration over an initially trained model is triggered when a specified number of new elements is available.

\subsection{Problem formulation} \label{subsec:problem_formulation}

The objective of our \textit{slicing for AI} problem is to optimize the allocation of resources to continuously train multiple AI models, $i \in \mathcal{I}$, corresponding to AI services, while collectively maximizing their performance and meeting the diverse services' requirements at each $t$ slot. To streamline presentation, we introduce the vector $\bm{a}^{(t)} = (L_{1}^{(t)},m_{1}^{(t)},\psi_{1}^{(t)},\lambda_{1}^{(t)},...,L_{I}^{(t)},m_{I}^{(t)},\psi_{I}^{(t)},\lambda_{I}^{(t)})$, that determines the allocation decision at the beginning of the $t$-th slot, where $I$ is the cardinality of $\mathcal{I}$. We define the performance function of the system as the normalized weighted summation of all AI model accuracies at the end of slot $t$:
\begin{equation}
\label{eq:AccSum}
 f^{(t)}(\bm{a}^{(t)})= \frac{\sum_{i}^{I} \alpha_{i} \cdot q_{i}^{(t)}(L_{i}^{(t)},m_{i}^{(t)})}{\sum_{i=1}^{I}\alpha_{i}},
\end{equation} 
where each AI model is associated with a fixed weighting coefficient $\alpha_{i}$ that determines the relative priority of the service. We remark that this type of problems is considered a bandit problem where the observed system performance feedback is only limited to the selected decision $\bm{a}^{(t)}$.

The challenge in the formulated problem is that $f^{(t)}(\bm{a})$ depends on $q_{i}^{(t)}$ for each $i \in \mathcal{I}$, which is initially unknown when the allocation decision is determined at beginning of $t$. Additionally, the accuracy $q_{i}^{(t)}$ of an AI model $i$ could adversarially change for the same pair of $L_{i}$ and $m_{i}$ according to some of the various previously mentioned factors as well as the availability of the training data (i.e., the observed system performance can unpredictably change from one time slot to another \cite{slivkins2022bandits}). In such an online system, the learner selects the allocation decision, $\bm{a}^{(t)}$, at the beginning of slot $t$ based on the preceding sequence of performance functions $\{f^{(\tau)}(\bm{a}^{(\tau)})\}_{\tau=1}^{t-1}$ (i.e., based on available information up to slot $t-1$).  Hence, in order to evaluate the performance of the online policy, we need to compare it against the system performance corresponding to the optimal allocation decision $\bm{a}^{\star} = (L_{1}^{\star},m_{1}^{\star},\psi_{1}^{\star},\lambda_{1}^{\star},...,L_{I}^{\star},m_{I}^{\star},\psi_{I}^{\star},\lambda_{I}^{\star})$, which is the solution to the following optimization problem with the objective to maximize the system performance:
\begin{align} 
\textbf{P: } \hspace{1em} &\max_{\bm{a}} \quad \sum_{t=1}^{T}{f^{(t)}(\bm{a})} \label{eq:P} \hspace{13em} \\
 & \textrm{subject to} \nonumber \\
 \quad &\sum_{i=1}^{I}{\psi_{i}} \leq{\Psi_\text{max}} \label{eq:Amax} \hspace{2em} \\
&\sum_{i=1}^{I}{\lambda_{i}}\leq{\Lambda_\text{max}} \label{eq:Rmax} \hspace{2em}\\
&D(L_{i},m_{i},\psi_{i},\lambda_{i})\leq D_\text{max$_{i}$}, \quad \forall{i} \in \mathcal{I}  \label{eq:Tmax} \\
& C(\psi_{i},\lambda_{i}) \leq C_\text{max$_{i}$}, \quad \hspace{1.15cm}   \forall{i} \in \mathcal{I}  \label{eq:Cmax} \\
 & L_\text{min$_{i}$} \leq L_{i} \leq L_\text{max$_{i}$},  \quad \hspace{0.82cm} \forall{i} \in \mathcal{I} \label{eq:L_limit}\\
 & m_\text{min$_{i}$} \leq m_{i} \leq m_\text{max$_{i}$},  \quad \hspace{0.6cm} \forall{i} \in \mathcal{I} \label{eq:m_limit}
\end{align}

The constraints in (\ref{eq:Amax}) and (\ref{eq:Rmax}) ensure that the summation of the allocated resources for all $i \in \mathcal{I}$ does not exceed the maximum available computing resources $\Psi_\text{max}$ and communication resources  $\Lambda_\text{max}$. The Constraints in (\ref{eq:Tmax}) and (\ref{eq:Cmax}) respectively represent the maximum allowable  latency $D_\text{max$_{i}$}$ and cost $C_\text{max$_{i}$}$ to train the AI model $i$. Herein, the learning cost at the $t$-th slot, $C(\psi_{i}^{(t)},\lambda_{i}^{(t)}$), is defined as the sum of the cost of allocated resources: 
 \begin{equation}
 \label{eq:Cost}
 C(\psi_{i}^{(t)},\lambda_{i}^{(t)}) = c_{\psi} \cdot \psi_{i}^{(t)} + c_{\lambda} \cdot \lambda_{i}^{(t)},
 \end{equation}
where $c_{\psi}$ is the cost per computing unit and $c_{\lambda}$ is the cost per unit of data rate. The allowable range of the data size (i.e., $[L_\text{min$_{i}$}, L_\text{max$_{i}$}]$) and the number of epochs (i.e., $[m_\text{min$_{i}$}, m_\text{max$_{i}$}]$) to train each AI model $i$ is respectively given by constraints (\ref{eq:L_limit}) and (\ref{eq:m_limit}).

It is clear that $\bm{a}^{\star}  = \arg \max_{\bm{a}\in \mathcal{A}}{\sum_{t=1}^{T}}f^{(t)}(\bm{a})$ is a hypothetical solution that can only be determined with hindsight (i.e., to devise such a decision, knowledge about the sequence of all future performance functions $\{f^{(t)}(\bm{a})\}_{t=1}^{T}$ is required) and we compare it with the online policy using the regret metric:
\begin{equation}
 R_{T} = \sum_{t=1}^{T}{f^{(t)}(\bm{a}^{\star})}-\sum_{t=1}^{T}{f^{(t)}(\bm{a}^{(t)})},
\end{equation}
which measures the loss accumulated across time as a result of not choosing the optimal action, $\bm{a}^\star$, at each slot $t$. The goal of an online policy is to achieve a sub-linear regret, $R_{T}=O(\sqrt{T})$, such that the loss decreases as $T$ grows, $\lim_{T\to\infty}\frac{R_{T}}{T} = 0$. A diminishing regret indicates that the online decisions are achieving a performance identical to that of the optimal action in hindsight.

The formulated problem depends on multiple integer variables (i.e., $(L_{i},m_{i},\psi_{i},\lambda_{i})$ $\forall i \in \mathcal{I}$) and hence it is a combinatorial optimization problem which can not be easily tackled with conventional optimization techniques. We prove the NP-hardness of the problem in the following Lemma.

\textbf{Lemma $1$:} \textit{
The {slicing for AI} problem formulated in \textbf{P} is NP-hard problem. }

\begin{IEEEproof} 
{
 The Knapsack problem \cite{NPhard}, which is a combinatorial optimization problem that is known to be NP-hard, is reducible to the formulated problem in \textbf{P}. The knapsack problem involves selecting a subset of items  $\mathcal{S} \subseteq \mathcal{N} = \{1,2,...,N\}$ from a set of $N$ items, such that each $n \in \mathcal{N}$ is characterized by a value $v_{n}$ and a weight $w_{n}$. The objective of the problem is to select $\mathcal{S}$ such that the total value $\sum_{n=1}^{N}{v_{n}}$ is maximized and while satisfying a maximum weight constraint $\sum_{n=1}^{N}{w_{n}} \le W$.  Correspondingly, the formulated problem in \textbf{P} aims at maximizing the system performance across all $t$ slots by determining the combination $\bm{a}^{\star} = (L_{1}^{\star},m_{1}^{\star},\psi_{1}^{\star},\lambda_{1}^{\star},...,L_{I}^{\star},m_{I}^{\star},\psi_{I}^{\star},\lambda_{I}^{\star})$ from the set of combinations $\mathcal{A} = \{\bm{a}_{1}, \bm{a}_{2},...\}$. Hence, each combination $\bm{a} \in \mathcal{A}$ representing an allocation decision can be mapped to the considered subset of items, $\mathcal{S}$, in the knapsack problem. The resulting inference accuracy of an allocation decision corresponding to each AI model can be mapped to the value variable in the knapsack problem, and the  
computing resources, communication resources, learning latency, and learning cost variables corresponding to each AI model can be mapped to the weight variable in the knapsack problem. This shows the formulated problem in \textbf{P} is a special case of the knapsack problem proving our Lemma that the  formulated problem is NP-hard.   
}
\end{IEEEproof} 


\section{Proposed Solutions of the slicing for AI problem\label{sec:Solution}}

In this section, we present an online learning framework to solve the formulated problem in \textbf{P}. We firstly propose Online Learning for Slicing (OLS); a basic online learning solution to address the formulated problem. We then present two variants of the solution to optimize the decision space; Online Learning for Slicing with Super Actions (OLS-SA), and Online Learning for Slicing with Reduced Super Actions (OLS-RSA). Particularly, the algorithms are composed of two phases: 1) The pre-learning phase, which involves pre-processing steps prior to the learning, 2) The learning phase based on the Exponential-weight algorithm for Exploration and Exploitation (EXP3) in \cite{OL12}. EXP3 is an adversarial bandit algorithm that does not require assumptions about the distribution of the environment's behavior, which perfectly fits the expected behavior of AI models. We highlight that each algorithm builds upon the pre-learning phase of the other. We present the details of the three algorithms in Sec. \ref{sec:OL}. Moreover, we propose a biased subset selection from the original decision space based on previous knowledge to improve the learning speed. We present two alternative approaches of considering the selected subset; Strictly Biased Selection (SBS), and Gaussian Biased Selection (GBS). The details of each approach are presented in Sec. \ref{sec:biased}

\begin{figure}[t!]
	\centering
		\scalebox{1.44}{\includegraphics[width=0.33 \textwidth]{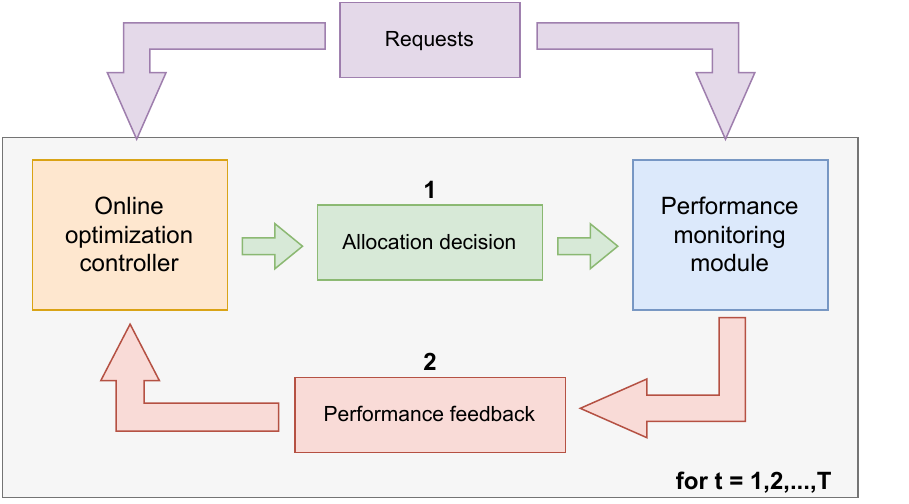}}
	\caption{The online learning framework, representing the interaction between the online optimization controller and a performance monitoring module}
	\label{fig:OL_framework}
\end{figure}

Figure \ref{fig:OL_framework}, represents the structure of the online learning framework. At each time slot $t$ the online optimization controller determines an allocation decision, $\bm{a}^{(t)}$, for the admitted AI model deployment requests. A performance monitoring module monitors the performance of each AI model based on the allocation decision and sends the system performance feedback, $f^{(t)}(\bm{a}^{(t)})$, to the online optimization controller, based on which the allocation decision at the next time step, $t+1$, is made.  



\subsection{Online Learning Solution} \label{sec:OL}
Now we will present the different variants of our online learning solution.
\

\subsubsection{Online Learning for Slicing}
\hfill


\begin{algorithm}
\caption{Online Learning for Slicing (OLS)}
\label{alg:OLSAI}
\begin{algorithmic}[1]
\State{\textbf{Input: }$c_{\psi}$, $c_{\lambda}$, $\Psi_\text{max}$, $\Lambda_\text{max}$, $D_\text{max$_{i}$}$, $C_\text{max$_{i}$}$, $L_\text{min$_{i}$}$, $L_\text{max$_{i}$}$, $m_\text{min$_{i}$}$, $m_\text{max$_{i}$}$} 
\State {\textbf{Initialize:} $\eta \in (0,1)$, $\bm{w}^{(1)} = (1/J, ..., 1/J)$ } \label{lst:line:OLSAIstart}
\State {Determine a set $\mathcal{O}$ of all possible hyper-parameters combinations for each AI model $i \in \mathcal{I}$} 
\State {Determine a set $\mathcal{S}$ of all possible resources combinations for each AI model $i \in \mathcal{I}$}
\For{each $\bm{s} \in \mathcal{S}$}
\If{constraints (\ref{eq:Amax}), (\ref{eq:Rmax}), and (\ref{eq:Cmax}) are satisfied}
 \State{store $\bm{s}$ in feasible combinations set $\mathcal{\tilde{S}}$ }
 \EndIf
\EndFor \label{lst:line:OLSAIend}
\For{each $\bm{o} \in \mathcal{O}$}
\For{each $\bm{s} \in \mathcal{\tilde{S}}$}
\If{constraint (\ref{eq:Tmax}) is satisfied}
\State\multiline{store the combination in the set of feasible actions $\mathcal{A}_\text{OLS}$ }
\EndIf
\EndFor 
\EndFor
\For{t = 1,2,...,T}  \label{lst:line:EXP3start}
\State {choose  action $\bm{a}^{(t)}$ from $\mathcal{A}_\text{OLS}$ according to $\bm{{w}^{(t)}}$}
\State {receive loss, $y_{j^{(t)}}^{(t)} \in [0,1]$, such that $j^{(t)}$ $\widehat{=}$ $\bm{a}^{(t)}$}
\State{\textbf{update}}
\State{$\qquad \tilde{w}_{j^{(t)}} = w_{j^{(t)}}^{(t)} e^{-\eta y_{j^{(t)}}^{(t)}/w_{j^{(t)}}^{(t)}}$}
\State{$\qquad$for $k \neq j^{(t)}, \tilde{w}_{k} = w_{k}^{(t)}$}
\State{$\qquad \forall{k}, w_{k}^{(t+1)} = \frac{\tilde{w}_{k}}{\sum_{k}\tilde{w}_{k}}$}
\EndFor \label{lst:line:EXP3end}
\State{\textbf{Output: }$L_{i}^{(t)}, m_{i}^{(t)},\psi_{i}^{(t)},\lambda_{i}^{(t)}$ $\forall i \in \mathcal{I}$}
\end{algorithmic}
\end{algorithm}

In OLS, we opt to set a base for the online learning solution to solve the formulated problem in \textbf{P}. The main steps of OLS are present in Algorithm \ref{alg:OLSAI}. The pre-learning phase, involves pre-processing steps to eliminate infeasible actions that violate the constraints of the problem. The first step is discretizing each decision variable (i.e., $L_{i}, m_{i}, \psi_{i}, \lambda_{i}$) corresponding to each AI model $i$ such that: $ L_{i} \in [L_\text{min$_{i}$},L_\text{max$_{i}$}],  m_{i} \in [m_\text{min$_{i}$},m_\text{max$_{i}$}],   \psi_{i} \in (0, \Psi_\text{max}]$, and $\lambda_{i} \in (0,\Lambda_\text{max}]$. A set $\mathcal{O}$ of all possible combinations of hyper-parameters (i.e., data size and number of epochs) for all $i \in \mathcal{I}$ is determined. Each element of the set $\mathcal{O}$ represents a vector of a single combination $\bm{o} = (L_{1},m_{1},L_{2},m_{2},...,L_{I},m_{I})$. A set $\mathcal{S}$ of all possible combinations of resources (i.e., computational and communication resources) is also determined, such that each element $\bm{s} = (\psi_{1}, \lambda_{1},\psi_{2}, \lambda_{2},...,\psi_{I}, \lambda_{I})$. As a first step in tackling infeasible decisions, the feasible combinations $\bm{s} \in \mathcal{S}$ that satisfy the constraints (\ref{eq:Amax}), (\ref{eq:Rmax}), and (\ref{eq:Cmax}) are stored in the set $\mathcal{\tilde{S}}$. Then, each combination of $\bm{o} \in \mathcal{O}$ with each $\bm{s} \in \mathcal{\tilde{S}}$ is checked for satisfying the delay constraint (\ref{eq:Tmax}). If the constraint is satisfied, the combination is denoted by the allocation decision vector, $\bm{a} = (L_{1},m_{1},\psi_{1},\lambda_{1},...,L_{I},m_{I},\psi_{I},\lambda_{I})$, and stored in the set of feasible decisions $\mathcal{A}_\text{OLS}$, which is fed to the learning phase presented in line \ref{lst:line:EXP3start} to line \ref{lst:line:EXP3end} of OLS. 

In the learning phase, at each slot $t$ the learner selects a decision $\bm{a}^{(t)}$ from the set $\mathcal{A}_\text{OLS}$ according to a probability vector $\bm{w}^{(t)}$. The vector $\bm{w}^{(t)}$ represents the probability of a decision being close to the optimal decision $\bm{a}^{\star}$. Conventionally, $\bm{w^{(1)}}$ is initialized as a uniform distribution, $\bm{w}^{(1)} = (1/J, ..., 1/J)$, such that $J = |\mathcal{A}_\text{OLS}|$. According to the allocation decision, $\bm{a}^{(t)}$, the learner suffers the cost vector $\bm{y}^{(t)} \in [0,1]^{J}$, which corresponds to the system's performance at each $\bm{a} \in \mathcal{A}_\text{OLS}$. As previously discussed in Sec. \ref{subsec:problem_formulation}, the system performance can only be observed at the selected decision $\bm{a}^{(t)}$. Accordingly, it is equal to zero for all the other un-selected decisions. Hence, we are interested in the loss $y_{j^{(t)}}^{(t)}$, such that $j^{(t)}$ corresponds to the index of the selected decision at the $t$-th time slot, $\bm{a}^{(t)}$:
\begin{equation}
    y_{j^{(t)}}^{(t)} = 1 - {f}^{(t)}(\bm{a}^{(t)}), \label{eq:loss}
\end{equation}
where this loss is integrated into updating the values of the probability vector $\bm{w}^{(t)}$. Accordingly, the learner constructs the vector $\bm{w}^{(t+1)}$ to use in the upcoming time step $t+1$. The probability vector update depends on the parameter $\eta \in (0,1)$ which is known as the learning rate. $\eta $ tunes the trade-off between exploration and exploitation. As time progresses, the online optimization controller learns to increase the probability of good actions and decrease the probability of bad actions. As time grows infinitely, the learner identifies the optimal allocation of resources, $\bm{a}^{\star}$, by utilizing the historical system performance feedback obtained from previous allocations. Since some of the decision variables are part of the constraints only and not the objective (i.e., $\lambda_{i}, \psi_{i}$), this gives a degree of freedom in terms of the trade-off between cost and learning latency of an AI model. As a result, we can have multiple actions that satisfy the objective in the same way, and thus leading to the same system performance. Subsequently, the learner might identify a subset of $K$ optimal allocation decisions, $\{\bm{a}_{1}^{\star},\bm{a}_{2}^{\star}, ..., \bm{a}_{K}^{\star}\} \subset \mathcal{A}_\text{OLS}$, which share the same hyper-parameters combination, $\bm{o} \in \mathcal{O}$, instead of identifying a single optimal decision. 


The time complexity of performing the pre-learning phase depends on: 1) Checking the feasibility of each $\bm{s} \in \mathcal{S}$, 2) Checking the feasibility of each $\bm{o} \in \mathcal{O}$ with each $\bm{s} \in \mathcal{\tilde{S}}$. However, $|\mathcal{O}||\mathcal{\tilde{S}}| \gg |\mathcal{S}|$, hence, the time complexity is $T_\text{PL}(\text{OLS}) = O(|\mathcal{O}||\mathcal{\tilde{S}}|)$. On the other hand, the time complexity of the learning phase at each slot $t$ depends on updating $\bm{w}^{(t)}$ $\forall$ 
 $\bm{a} \in \mathcal{A}_\text{OLS} $, hence the complexity of the learning phase is $T_\text{L}(\text{OLS}) = O(|\mathcal{A}_\text{OLS}|T)$. Therefore the cumulative time complexity is $T_\text{cum}(\text{OLS}) = T_\text{PL}(\text{OLS})+T_\text{L}(\text{OLS})$.

\hfill
\subsubsection{Online Learning for Slicing with Super Actions}
\hfill

In the OLS-SA algorithm we aim for a method to combine the decisions that satisfy the objective in the same way to obtain a compact structure of the decision space. OLS-SA extends the pre-learning phase of the OLS algorithm to construct what we define as super actions. A super action represents a group/set of sub actions/decisions that share the same combination $\bm{o} \in \mathcal{O}$. Algorithm \ref{alg:EOLSAI} demonstrates the main steps of OLS-SA.
\begin{algorithm}
\caption{Online Learning for Slicing with Super Actions (OLS-SA)}
\label{alg:EOLSAI}
\begin{algorithmic}[1] 
\State{\textbf{Input: }$c_{\psi}$, $c_{\lambda}$, $\Psi_\text{max}$, $\Lambda_\text{max}$, $D_\text{max$_{i}$}$, $C_\text{max$_{i}$}$, $L_\text{min$_{i}$}$, $L_\text{max$_{i}$}$, $m_\text{min$_{i}$}$, $m_\text{max$_{i}$}$} 
\State{Run algorithm \ref{alg:OLSAI} from line \ref{lst:line:OLSAIstart} to line \ref{lst:line:OLSAIend}} 
\For{each $\bm{o} \in \mathcal{O}$}
\State{construct an empty super action set $\mathcal{B}_\text{SA}$ } \label{lst:line:EOLSAIstart}
\For{each $\bm{s} \in \mathcal{\tilde{S}}$}
\If{constraint (\ref{eq:Tmax}) is satisfied}
\State{store the combination in $\mathcal{B}_\text{SA}$}
\EndIf
\EndFor \label{lst:line:EOLSAIend}
\If{$\mathcal{B}_\text{SA}$ consists feasible sub actions}
\State\multiline{store $\mathcal{B}_\text{SA}$ in the family of sets of super actions $\mathcal{A}_\text{OLS-SA}$}
\EndIf
\EndFor
\State{Run the learning phase of Algorithm \ref{alg:OLSAI} from line \ref{lst:line:EXP3start} until line \ref{lst:line:EXP3end} to determine the selected super action at the $t$-th slot $\mathcal{B}_\text{SA}^{(t)}$}
\State{Choose a sub action $\bm{a}^{(t)}$ from $\mathcal{B}_\text{SA}^{(t)}$ at random}\label{lst:line:solvingEnd}
\State{\textbf{Output: }$L_{i}^{(t)}, m_{i}^{(t)},\psi_{i}^{(t)},\lambda_{i}^{(t)}$ $\forall i \in \mathcal{I}$}
\end{algorithmic}
\end{algorithm}

In Algorithm \ref{alg:EOLSAI}, the pre-learning phase starts by determining the set of feasible resources $\mathcal{\tilde{S}}$ in terms of satisfying cost and resources constraints. Then, for each $\bm{o} \in \mathcal{O}$, an empty set, $\mathcal{B}_\text{SA}$, which denotes a super action is constructed. Each combination $\bm{o} \in \mathcal{O}$ with $\bm{s} \in \mathcal{\tilde{S}}$ is checked for feasibility in terms of satisfying the learning latency constraint. Feasible combinations are then stored as the vector $\bm{a} = (L_{1},m_{1},\psi_{1},\lambda_{1},...,L_{I},m_{I},\psi_{I},\lambda_{I})$ in the set representing the super action $\mathcal{B}_\text{SA}$. Hence, at the end of checking the feasibility of each $\bm{o} \in \mathcal{O}$, if the set $\mathcal{B}_\text{SA}$ is not empty, this indicates that there exists feasible sub-actions sharing the same hyper-parameter combination. Consequently, $\mathcal{B}_\text{SA}$ is stored in a set that corresponds to the family of sets of super actions denoted as $\mathcal{A}_\text{OLS-SA}$. After completing the pre-learning phase, the set $\mathcal{A}_\text{OLS-SA}$ is fed to the learning phase such that at each slot $t$ the selected super action $\mathcal{B}_\text{SA}^{(t)}$ is determined as explained in the OLS algorithm. Since all sub actions $a^{(t)} \in \mathcal{B}_\text{SA}^{(t)}$ satisfy the constraints and lead to the same system performance, therefore, the selection of a specific allocation decision $a^{(t)} \in \mathcal{B}_\text{SA}^{(t)}$ at the $t$-th slot is performed randomly. At each slot $t$ the probability vector $\bm{w}^{(t)}$ is updated such that $j^{(t)}$ corresponds to $\mathcal{B}_\text{SA}^{(t)}$. As time progresses, the learner identifies the set corresponding to the optimal super action which includes the $K$ optimal allocation decisions as identified by the OLS algorithm, $\mathcal{B}_\text{SA}^{\star} = \{\bm{a}_{1}^{\star},\bm{a}_{2}^{\star}, ..., \bm{a}_{K}^{\star}\}$.   

The notion of super actions optimizes the size of the decision space presented to the learning phase of OLS-SA and consequently optimizes the cumulative time complexity. The pre-learning phase depends on: 1) Identifying feasible decisions, 2) Combining similar decisions into super actions. However, combining similar actions does not entail a significant time complexity overhead, hence the time complexity of the pre-learning phase is $T_\text{PL}(\text{OLS-SA}) = O(|\mathcal{O}||\mathcal{\tilde{S}}|)$. The time complexity of the learning phase is $T_\text{L}(\text{OLS-SA}) = O(|\mathcal{A}_\text{OLS-SA}|T)$, with $T_\text{cum}(\text{OLS-SA}) = T_\text{PL}(\text{OLS-SA}) + T_\text{L}(\text{OLS-SA})$. The cardinality $|\mathcal{A}_\text{OLS-SA}|$ is equal to the number of super action elements inside the set. Since, $|\mathcal{A}_\text{OLS-SA}| < |\mathcal{A}_\text{OLS}|$, therefore, $T_\text{cum}(\text{OLS-SA}) < T_\text{cum}(\text{OLS})$.

\hfill
\subsubsection{Online Learning for Slicing with Reduced Super Actions}
\hfill

\begin{algorithm}
\caption{Online Learning for Slicing with Reduced Super Actions (OLS-RSA)}
\label{alg:OOLSAI}
\begin{algorithmic}[1]
\State{\textbf{Input: }$c_{\psi}$, $c_{\lambda}$, $\Psi_\text{max}$, $\Lambda_\text{max}$, $D_\text{max$_{i}$}$, $C_\text{max$_{i}$}$, $L_\text{min$_{i}$}$, $L_\text{max$_{i}$}$, $m_\text{min$_{i}$}$, $m_\text{max$_{i}$}$} 
\State\multiline{\textbf{Initialize} an empty set $\mathcal{A}_\text{OLS-RSA}$ to store candidates of the optimal super action}
\State{Run algorithm \ref{alg:OLSAI} from line \ref{lst:line:OLSAIstart} to line \ref{lst:line:OLSAIend}}
\For{each $o \in \mathcal{O}$}
\State{Run Algorithm \ref{alg:EOLSAI} form line \ref{lst:line:EOLSAIstart} to line \ref{lst:line:EOLSAIend}}
\If{$\mathcal{B}_\text{SA}$ consists feasible sub actions}
\State\multiline{store $\mathcal{B}_\text{SA}$ in the family of sets of super actions $\mathcal{A}_\text{OLS-RSA}$ if it is empty}
\State{copy $\mathcal{A}_\text{OLS-RSA}$ into the set $\mathcal{\tilde{A}}$}
\For{each $ \mathcal{\tilde{B}}_\text{SA} \in \mathcal{\tilde{A}}$}
\State\multiline{check for candidacy of the super action $\mathcal{B}_\text{SA}$ over $\mathcal{\tilde{B}}_\text{SA}$ according to criteria 1}
\State\multiline{store the super actions $\mathcal{\tilde{B}}_\text{SA}$ with overtook candidacy in the set $\mathcal{R}$}
\State\multiline{check for candidacy of $\mathcal{B}_\text{SA}$ according to criteria 2}
\EndFor
\If{$\mathcal{B}_\text{SA}$ $isCandidate$ }
\State{store in the set $\mathcal{\tilde{A}}$}
\State{\textbf{update}: $\mathcal{A}_\text{OLS-RSA} =  \mathcal{\tilde{A}} - \mathcal{R}$}
\EndIf
\EndIf
\EndFor
\State{Run the learning phase of Algorithm \ref{alg:OLSAI} from line \ref{lst:line:EXP3start} until line \ref{lst:line:EXP3end} to determine the selected super action at the $t$-th slot $\mathcal{B}_\text{SA}^{(t)}$}
\State{Choose a sub action $\bm{a}^{(t)}$ from $\mathcal{B}_\text{SA}^{(t)}$ at random}
\State{\textbf{Output: }$L_{i}^{(t)}, m_{i}^{(t)},\psi_{i}^{(t)},\lambda_{i}^{(t)}$ $\forall i \in \mathcal{I}$}
\end{algorithmic}
\end{algorithm}

The OLS-RSA algorithm is proposed based on the principle of Machine Learning (ML) that increasing the number of epochs and data size used for training AI models results in higher accuracy. In OLS-RSA, we opt for a method to identify candidates of the optimal super action for reducing the size of the decision space. We define as the candidates of an optimal super action the super actions with hyper-parameters combinations that consists of the highest number of epochs and/or training data size. For example, we can have the three candidates of the optimal super action $\mathcal{B}_\text{SA$_{1}$}$, $ \mathcal{B}_\text{SA$_{2}$}$, and $ \mathcal{B}_\text{SA$_{3}$}$ respectively corresponding to the hyper-parameters combinations $(L_{1,1},m_{1,1},L_{2,1},m_{2,1})$, $(L_{1,2},m_{1,2},L_{2,2},m_{2,2})$, and $(L_{1,3},m_{1,3},L_{2,3},m_{2,3})$, where $L_{1,1} > L_{1,2} > L_{1,3}$, $m_{1,1} < m_{1,2} < m_{1,3}$, $L_{2,1} = L_{2,2} = L_{2,3}$, and $m_{2,1} = m_{2,2} = m_{2,3}$. That is, the three super actions exhibit a trade-off between a higher number of epochs against a higher data size corresponding to AI model $1$. This motivates us to propose the OLS-RSA algorithm which extends the pre-learning phase of the OLS-SA algorithm to identify the super actions that possesses such a trade-off, while optimizing the size of the decision space. This involves an additional step of eliminating non-candidates and hence the name "Reduced Super Actions".

As a first step in Algorithm \ref{alg:OOLSAI}, the set of feasible resources combinations $\mathcal{\tilde{S}}$ is determined. 
Then, for every $\bm{o} \in \mathcal{O}$ the feasible combinations with $\bm{s} \in \mathcal{\tilde{S}}$ are stored in the super action set $\mathcal{B}_\text{SA}$. The first feasible super action $\mathcal{B}_\text{SA}$ is stored in an initially empty decision space $\mathcal{A}_\text{OLS-RSA}$, which represents a family of candidate optimal super action sets. In the succeeding iterations, $\mathcal{A}_\text{OLS-RSA}$ is copied into an intermediary set $\mathcal{\tilde{A}}$ to start the process of optimal super action candidacy checking. Candidacy is determined based on two different criteria: 1) Criteria 1: Candidacy of the super action $\mathcal{B}_\text{SA}$ over any of the super actions $\mathcal{\tilde{B}}_\text{SA} \in \mathcal{\tilde{A}}$ previously stored in the decision space, 2) Criteria 2: Candidacy in addition to the existing super actions in the set $\mathcal{\tilde{A}}$. The first criteria checks whether the shared hyper-parameters combination, $(L_{1},m_{1},...,L_{I},m_{I})$, corresponding to the super action $\mathcal{B}_\text{SA}$ could overtake the candidacy of another hyper-parameter combination, $(\tilde{L_{1}},\tilde{m_{1}},...,\tilde{L_{I}},\tilde{m_{I}})$, corresponding to $\mathcal{\tilde{B}}_\text{SA} \in \mathcal{\tilde{A}}$. If $L_{i} \geq \tilde{L_{i}}$ and $m_{i} \geq \tilde{m_{i}}$ $\forall i \in \mathcal{I}$, then the super action $\mathcal{B}_\text{SA}$ overtakes the candidacy from $\mathcal{\tilde{B}}_\text{SA}$. Subsequently, the action $\mathcal{B}_\text{SA}$ is marked as a candidate by the flag $isCandidate$, while all the super actions $\mathcal{\tilde{B}}_\text{SA}$ with overtook candidacy are added to the set $\mathcal{R}$ for elimination. The second criteria checks whether the action $\mathcal{B}_\text{SA}$ is a new additional candidate. Specifically, if $L_{i} > \tilde{L_{i}}$ \&  $m_{i} < \tilde{m_{i}}$ or $L_{i} \leq \tilde{L_{i}}$ \&  $m_{i} > \tilde{m_{i}}$, for any $i \in \mathcal{I}$ then $\mathcal{B}_\text{SA}$ is marked as a candidate by the flag $isCandidate$. After that, a super action $\mathcal{B}_\text{SA}$ that is marked as $isCandidate$ is added to the set $\mathcal{\tilde{A}}$, followed by an elimination of the super actions with overtook candidacy by updating the decision space, $\mathcal{A}_\text{OLS-RSA} =  \mathcal{\tilde{A}} - \mathcal{R}$. Finally, the learning phase is executed as explained in the OLS-SA algorithm to obtain $\mathcal{B}_\text{SA}^{(t)}$ and $\bm{a}^{(t)}$, then solve for the optimization variables (i.e., $L_{i}, m_{i},\psi_{i},\lambda_{i}$ $\forall i \in \mathcal{I}$). As time progresses, the learner is able to identify the same optimal super action and accordingly the optimal allocation decisions identified by the OLS-SA and OLS algorithms, $\mathcal{B}_\text{SA}^{\star} = \{\bm{a}_{1}^{\star},\bm{a}_{2}^{\star}, ..., \bm{a}_{K}^{\star}\}$. 

The time complexity of the pre-learning phase $T_\text{PL}(\text{OLS-RSA})$ depends on three main steps: 1) Determining the feasible decisions, 2) Combining similar decisions into super actions, 3) Determining optimal super action candidates. The step  of candidacy checking depends on comparing every new feasible super action $\mathcal{B}_\text{SA}$ with candidates previously stored in the decision space as well as removing super actions with overtook candidacy. On the other hand, the time complexity of the learning phase is $T_\text{L}(\text{OLS-RSA}) = O(|\mathcal{A}_{3}|T)$, with $T_\text{cum}(\text{OLS-RSA}) = T_\text{PL}(\text{OLS-RSA}) + T_{L}(\text{OLS-RSA})$. The set $\mathcal{A}_\text{OLS-RSA}$ is a reduced version of $\mathcal{A}_\text{OLS-SA}$, with the cardinality $|\mathcal{A}_\text{OLS-RSA}| < |\mathcal{A}_\text{OLS-SA}|$ representing the number of candidates of the optimal super action. Hence, $T_\text{cum}(\text{OLS-RSA}) < T_\text{cum}(\text{OLS-SA}) < T_\text{cum}(\text{OLS})$.

\subsection{Biased action space subset selection} \label{sec:biased}

The learning phase of the algorithms, OLS, OLS-SA, and OLS-RSA, guarantees an upper regret bound to the accumulated loss during the $T$ time slots which is given by \cite[Corollary~4.2] {OL12}:
\begin{equation}
    R_{T} \leq \frac{\log(J)}{\eta} + \eta JT, \label{eq:regret_bound}
\end{equation}
Ideally, in a highly adversarial environment, we are interested in the optimal (i.e., smallest) upper regret bound. This is achieved at the optimal $\eta$ given by \cite{OL12}:
\begin{equation}
    \eta_\text{op} = \sqrt{\log(J)/(JT)} \label{optimal_eta}
\end{equation}

It is a common practice in the learning phase to initialize $\bm{w}_{(1)}$ as uniformly distributed among all actions/super actions in a decision space denoted as $\mathcal{A}$. However, if we can identify a subset of $\mathcal{A}$ within which the optimal action/super action is expected to lie based on prior knowledge, it makes sense to rethink the initial distribution of probabilities such that it biases this subset to optimize the upper regret bound. We assume that this subset can be identified based on a previous experience considering a similar request of resource allocation for training AI models. Hence, we put forth two alternative approaches of biasing this subset. We emphasize that these approaches concern the tuning of the probability vector at the initialization step of the presented online solution in Sec. \ref{sec:OL}.
\subsubsection{Strictly Biased Subset (SBS)}
\hfill

We consider a subset of actions with size $J' < J$ over which the initial probability is uniformly distributed, such that  $J = \mathcal{|A|}$. Consequently, the probability corresponding to each action within that subset is equal to $1/J'$ while it is given a value of zero for the rest of the actions in the decision space and hence the name "Strictly Biased Subset". Thus, from equation (\ref{eq:regret_bound}) we re-define the upper regret bound as:
\begin{equation}
     R'_{T} \leq \frac{\log(J')}{\eta} + \eta J'T, \label{eq:biased_regret_bound}
\end{equation}
therefore, the optimal learning rate $\eta_\text{op}'$ for a highly adversarial environment is re-defined as:
\begin{equation}
    \eta_\text{op}' = \sqrt{\log(J')/(J'T)} \label{biased_optimal_eta}
\end{equation}
Since, $J' < J$ we have $R'_{T} < R_{T}$ and $\eta_\text{op}' > \eta_\text{op}$,
where a smaller size of the subset $J'$ will result in a lower $R'_{T}$ and a larger $\eta_\text{op}'$.

\subsubsection{Gaussian Biased Subset (GBS)}
\hfill

\begin{figure}[t!]
	\centering
		\scalebox{1.4}{\includegraphics[width=0.33 \textwidth]{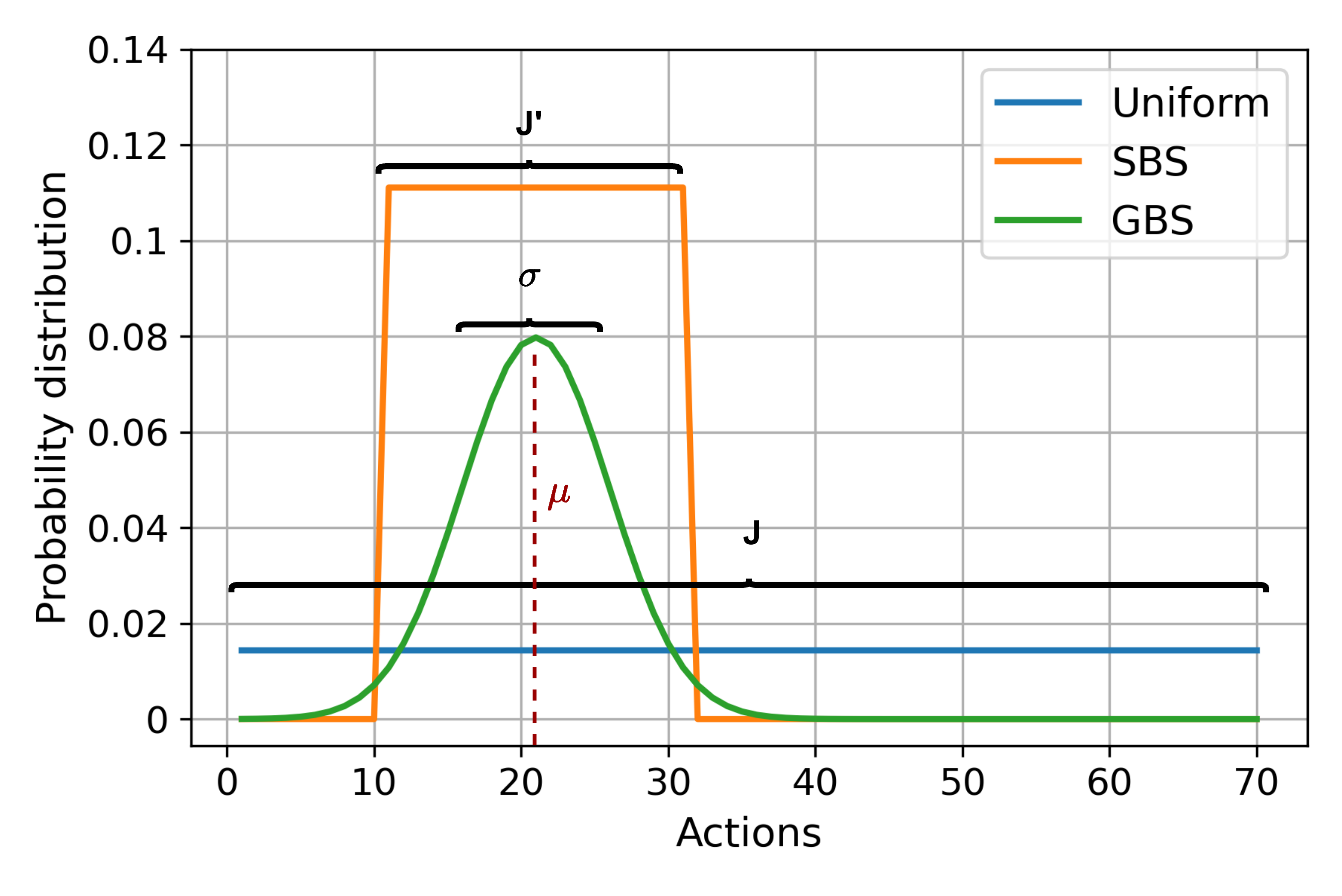}}
	\caption{A representation of different initial probability distribution schemes}
	\label{fig:weights_schemes}
\end{figure}

Although the subset can be identified based on previous knowledge, there is a possibility that it may not include the optimal action since distinct AI models may exhibit different performance behaviours. Therefore, instead of completely omitting the possibility of choosing an action outside the identified subset, we propose a biased selection based on Gaussian distribution.
A Gaussian distribution is characterized by its mean $\mu$ and standard deviation $\sigma$ (see Figure \ref{fig:weights_schemes}). A smaller $\sigma$ around the optimal action/super action corresponds to a probability density that is more concentrated around a smaller subset of the decision space. The proximity of $\mu$ to the optimal action determines the error in identifying the subset which it belongs to.

\section{Performance Evaluation \label{sec:Simulation}}
In this section, we will present the considered experimental setup, the conducted experiments, and results to evaluate the performance of the proposed solutions.


\subsection{Experiment Setup}

We perform our experiment considering the Deep Learning (DL) model for a mobile health application and the dataset presented in \cite{Model}. For the sake of performance evaluation, we derive an exponential regression model to simulate the system's behavior in terms of inference accuracy of the considered DL model. To obtain a regression model, we train the DL model using different data sizes and number of epochs. We then observe the corresponding inference accuracy on some test data by averaging the results over 10 experiments. The considered DL model consists of 11 layers and is trained on 245,921 samples of the associated dataset, which consists of 13 classes representing different physical activities. Our regression model captures the dependence of the inference accuracy, $q_{i}$, on the number of epochs, $m_{i}^{(t)}$, and the percentage of the training data, $l_{i}^{(t)}$. The percentage of the data size, $l_{i}^{(t)}$, is considered with respect to the 245,921 training samples (i.e., $l_{i}^{(t)} = (L_{i}^{(t)}/245,921) \times 100$). We use the following exponential model to regress the relation between the three variables: 
\begin{equation}
    q_{i}(l_{i}^{(t)},m_{i}^{(t)}) = \frac{1}{100}( g_{1}e^{g_{2}l_{i}^{(t)}}+ g_{3}e^{g_{4}m_{i}^{(t)}}+g_{5}e^{g_{6}m_{i}^{(t)}}) \label{eq:model_accuracy}
\end{equation}

The coefficients corresponding to DL model 1 presented in Table \ref{tab:services} is the result of performing the regression, with a root-mean square error (RMSE) of $5.663$ and an R-square measure of $0.9379$. The RMSE and the R-square measures indicates that the model captures the relation well, and hence is a good fit. We replicate the obtained exponential model of DL model 1 by adjusting the values of the coefficients and obtain regression models to represent three more DL models (see Table \ref{tab:services}).  We highlight that an adversarial behavior can be represented by a changing exponential model, (i.e., $q_{i}^{(t)}(l_{i}^{(t)},m_{i}^{(t)})$), or coefficients, (i.e., $g_{1}^{(t)}$, $g_{2}^{(t)}$, $g_{3}^{(t)}$, $g_{4}^{(t)}$, $g_{5}^{(t)}$, 
 $g_{6}^{(t)}$), at each time slot according to the distribution and quality of the data. However, for simplicity and without loss of generality, we consider data with fixed distribution and quality. Hence, we fix the model (\ref{eq:model_accuracy}) and its corresponding coefficients for all the considered DL models throughout the entire time horizon. 
{ \small
\renewcommand{\arraystretch}{1.5}
\begin{table}[t!]
	\centering
\caption{Coefficients of the considered DL models}
	  \label{tab:services}
\begin{tabular}{|c|c|} 
			\hline 
\textbf{Coefficients} & \textbf{($g_{1}$, $g_{2}$, $g_{3}$, $g_{4}$, $g_{5}$, 
 $g_{6}$)} 
\\
\hline
\textbf{DL model 1} & (-60, -0.03109, 96.98, 0.0006553, -120, -0.8355)
  \\
\hline
\textbf{DL model 2} &(-48, -0.03, 98.5, 0.001, -97, -0.5)
  \\
\hline
\textbf{DL model 3} & (-40, -0.04, 97, 0.002, -110, -0.6)
  \\
\hline
\textbf{DL model 4} & (-38, -0.04, 95, 0.0015, -100, -0.64)
  \\
\hline
\end{tabular} 
\end{table}
 }
 
After obtaining the regression models, we consider the DL model deployment requests coming as a tuple represented by: $(\alpha, C_\text{max}, D_\text{max} , l_\text{min}, l_\text{max}, m_\text{min}, m_\text{max})$, where $C_\text{max}$ is the maximum budget in dollars (\$), $D_\text{max}$ is the maximum latency in minutes (mins), $[l_\text{min}, l_\text{max}]$ is the range of data size in percentage. The resource allocation problem is solved by the online optimization controller considering the availability and cost of the resources determined by the network operator. The available resources and cost is represented by the tuple: $(\Psi_\text{max},\phi,\Lambda_\text{max},c_{\psi},c_{\lambda})$. $\Psi_\text{max}$ is the available units of CPU resources in gigahertz (GHz), and $\Lambda_\text{max}$ is the available bandwidth in batches per second (batches/sec), where 10000 samples of data constitute a batch. $c_{\psi}$ and $c_{\lambda}$ represent the cost of a computing unit and a unit rate in dollars respectively. We consider the simulation parameters presented in Table \ref{tab:simulation_parameters} and assume a negligible channel access delay (i.e., $\epsilon = 0$).
 { \small
\renewcommand{\arraystretch}{1.5}
\begin{table}[t!]
	\centering
\caption{Simulation parameters}
	  \label{tab:simulation_parameters}
\begin{tabular}{|c|c|c|c|c|c|} 
\hline 
\multicolumn{6}{|c|}{\textbf{Service Requirements}}\\
\hline
\multicolumn{2}{|c|}{\textbf{Request}} & \multicolumn{4}{|c|}{$(\alpha, C_\text{max}, D_\text{max} , l_\text{min}, l_\text{max}, m_\text{min}, m_\text{max})$ } \\
\hline
\multicolumn{2}{|c|}{\textbf{DL model 1}} & \multicolumn{4}{|c|}{(1, 0.46, 3.70, 25, 100, 2, 10)}\\
\hline
\multicolumn{2}{|c|}{\textbf{DL model 2}} & \multicolumn{4}{|c|}{(1, 0.36, 4.50, 25, 100, 2, 10)}\\
\hline
\multicolumn{2}{|c|}{\textbf{DL model 3}} & \multicolumn{4}{|c|}{(1, 0.36, 2.43, 25, 100, 2, 10)}\\
\hline
\multicolumn{2}{|c|}{\textbf{DL model 4}} & \multicolumn{4}{|c|}{(1, 0.36, 5.3, 20, 100, 3, 10)}\\
\hline
\multicolumn{6}{|c|}{\textbf{Available Resources \& Cost}}\\
\hline
\multicolumn{3}{|c|}{$(\Psi_\text{max},\phi,\Lambda_\text{max},c_{\psi},c_{\lambda})$} & \multicolumn{3}{|c|}{(3.7, 350000, 5, 0.2, 0.02)}\\
\hline
\end{tabular} 

\end{table}
 }

 To evaluate the performance of our proposed solution, we compare it against two benchmarks. The first is Optimal Allocation (OA) where we compare how well the learning performs against the optimal system performance at each time slot $f^{(t)}(a^{\star})$. The optimal decision $a^{\star}$ is determined by solving the formulated problem in \textbf{P} using a greedy search through the discretized feasible decision space $\mathcal{A}$. As previously explained in Sec. \ref{sec:OL}, there may be more than a single optimal allocation decision, $\{\bm{a}_{1}^{\star},\bm{a}_{2}^{\star}, ..., \bm{a}_{K}^{\star}\}$. The second benchmark is fixed allocation (FA), where the network operator decides on the amount of computing and communication resources to be allocated to any admitted slice. This allocation is fixed throughout the lifetime of the slices and does not guarantee optimality. For the ease of simulation, we consider a small scale scenario.

\subsection{Learning rate analysis}
\begin{figure*}
     \centering
     \begin{subfigure}[b]{0.325\textwidth}
         \centering
         \includegraphics[width=\textwidth]{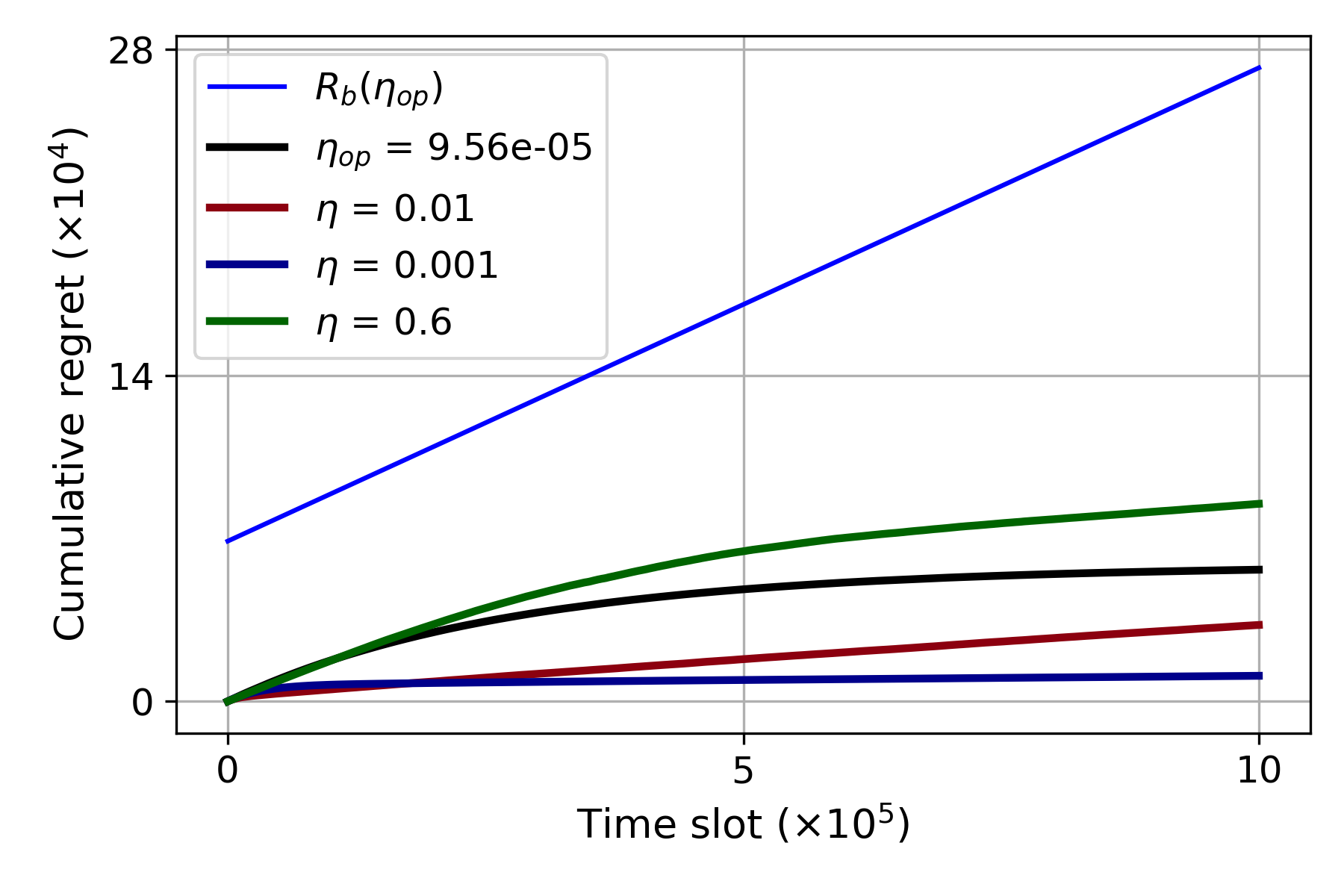}
         \caption{}
         \label{fig:cumulative_regret}
     \end{subfigure}
     \begin{subfigure}[b]{0.325\textwidth}
         \centering
         \includegraphics[width=\textwidth]{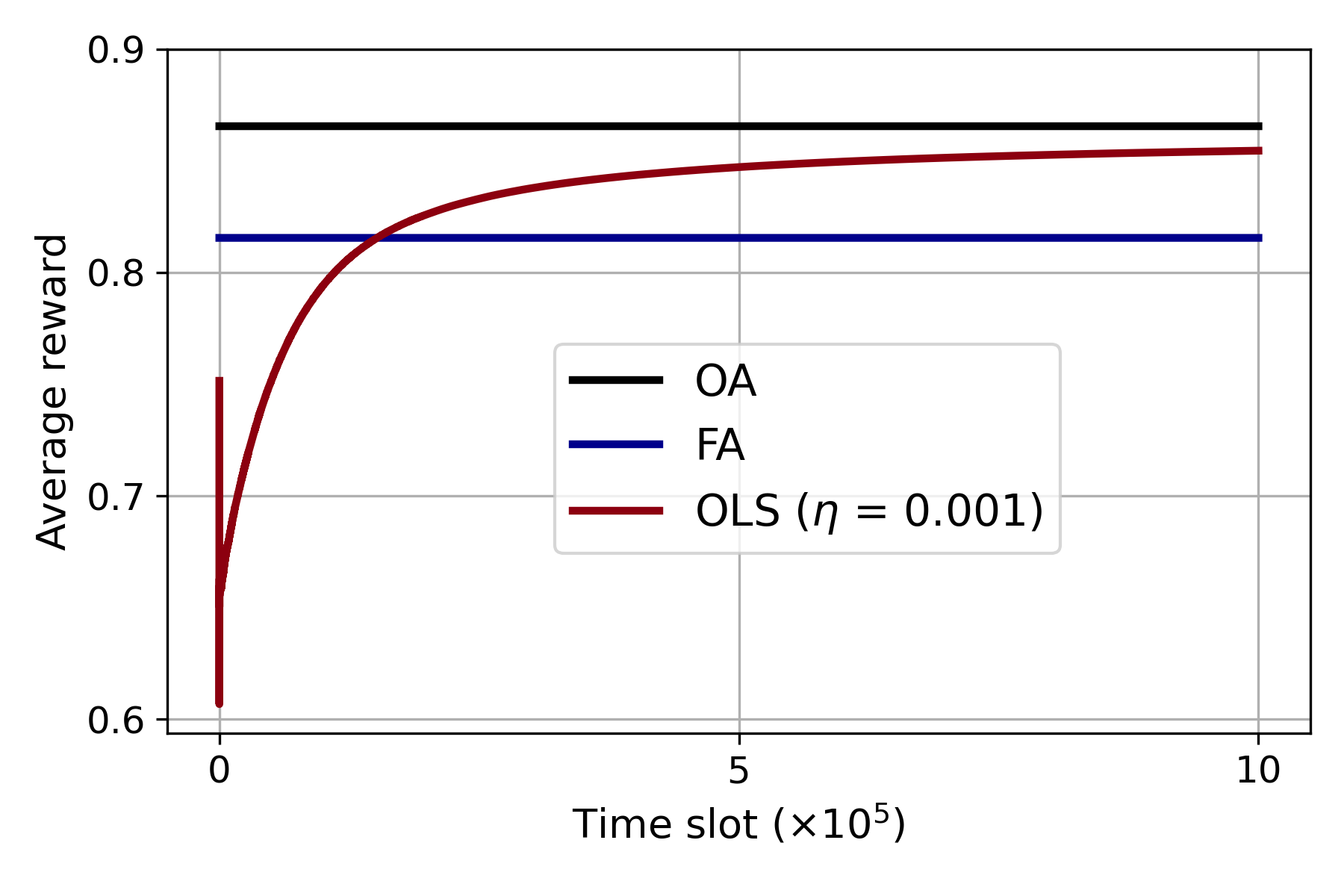}
         \caption{}
         \label{fig:benchmarks_comparison}
     \end{subfigure}
          \begin{subfigure}[b]{0.325\textwidth}
         \centering
         \includegraphics[width=\textwidth]{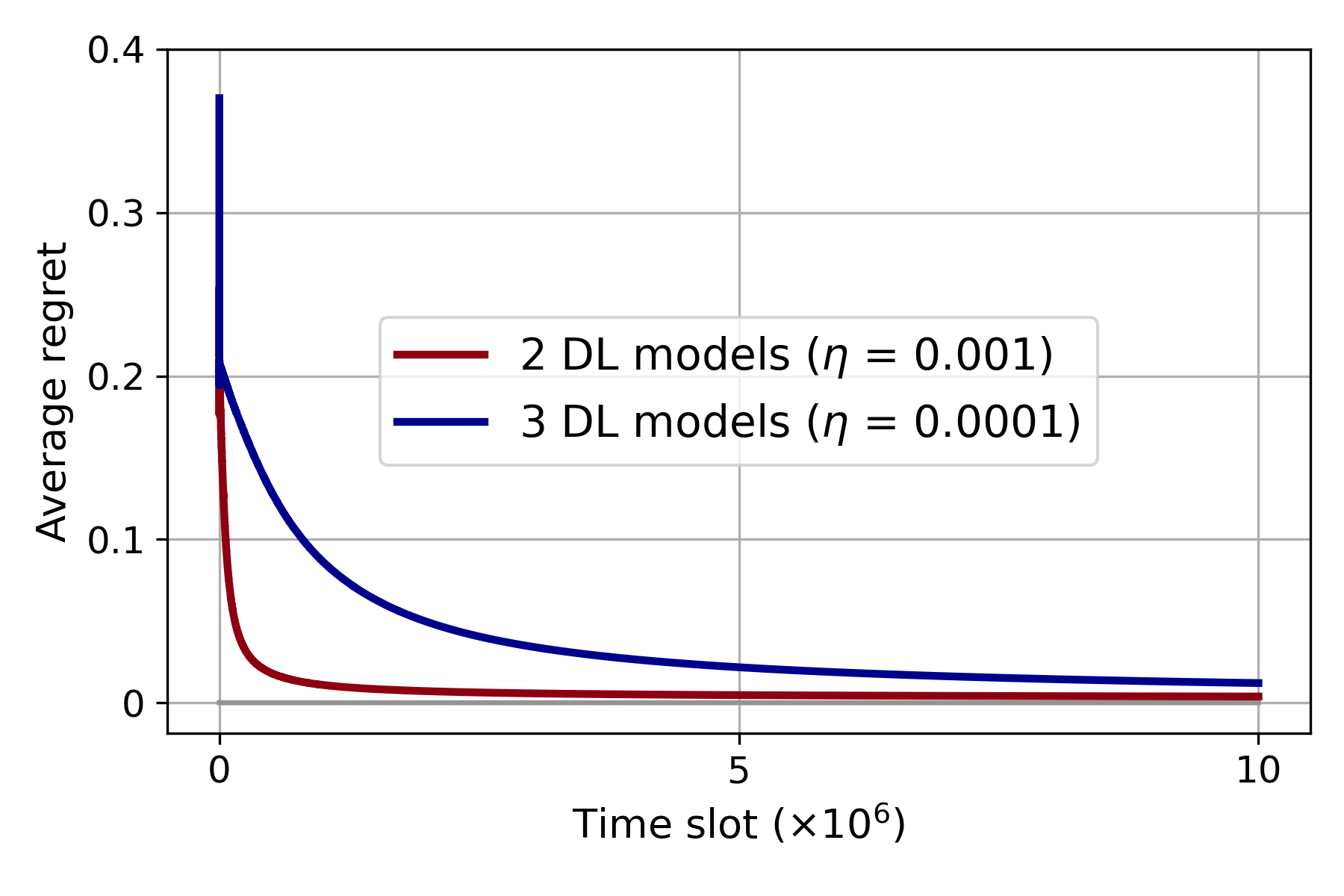}
         \caption{}
         \label{fig:setups_comparison}
     \end{subfigure}
     \caption{a) Cumulative regret curve using different values for $\eta$ compared against the regret bound at $\eta_\text{op}$, b) Average reward curve of OLS against the benchmarks using the best observable $\eta$, c) Average regret curve considering different numbers of DL model deployment requests using the best observed $\eta$}
     \label{fig:Experiment1}
\end{figure*}

We study the impact of the learning rate $\eta$ on the performance of a learner, the best observable $\eta$, and the impact of changing the environment on the behavior of $\eta$. 
we run an experiment considering DL model 1, DL model 2 (see Table \ref{tab:services}), and the available resources presented in Table \ref{tab:simulation_parameters}. The action space is discretized according to the following: 
\begin{itemize}
    \item $l_{{i}} \in \{25, 50, 100\}$ $\forall i \in \mathcal{I}$
    \item $m_{i} \in \{2,5,10\}$ $\forall i \in \mathcal{I}$
    \item $\psi_{i} \in \{1.5,1.8,2.2\}$ $\forall i \in \mathcal{I}$
    \item $\lambda_{i} \in \{1,2,3\}$ $\forall i \in \mathcal{I}$
\end{itemize}
We compare the performance of OLS algorithm, in terms of the cumulative regret (i.e., $R_{T}$), using $\eta = \eta_\text{op}$ obtained from equation (\ref{optimal_eta}), against its performance using different values for $\eta$. 
The results in Figure  \ref{fig:Experiment1}-\subref{fig:cumulative_regret} shows that although the optimal $\eta$ guarantees the lowest regret bound (i.e., the regret bound at $(\eta_\text{op})$) for a highly adversarial environment, we could observe a lower cumulative regret using another value of $\eta$ in a less adversarial environment. We observe that $\eta = 0.001$ exhibits the lowest cumulative regret and hence we consider it as the best observable learning rate for the simulated environment setup. We proceed with comparing the performance of OLS, in terms of the average reward (i.e., the average system performance; $\sum_{t=1}^{T}{f^{(t)}(\bm{a}^{(t)})}/T$), against the benchmark schemes OA and FA using $\eta = 0.001$ as shown in Figure \ref{fig:Experiment1}-\subref{fig:benchmarks_comparison}. We assume an FA scheme which allocates the following fixed resources: $l_{{i}} = 50$, $m_{i} = 5$, $\Psi_{i} = 1.5$, $\lambda_{i} = 2$, $\forall i \in \mathcal{I}$. It is clear that unlike the FA scheme, OLS is able to update the slices/allocation decision at each time step according to the observed system performance until the average reward converges to the optimal system performance as given by OA. 

In Figure \ref{fig:Experiment1}-\subref{fig:setups_comparison}, we study the behavior of another setup considering the first three DL models in Table \ref{tab:simulation_parameters} and $\Psi_\text{max} = 5.2$. We compare the performance of the learner while considering only two DL model deployment requests, against its performance considering an additional DL model deployment request in terms of the average regret (i.e., $R_{T}/T$). We identify that the best observable learning rate changes according to the environment setup. We particularly note that by accepting another DL model deployment request and up-scaling the available resources, the size of the decision space increases, which demands a lower value for $\eta$. It is also observed that learning on a smaller decision space results in a smaller area under the average regret curve (i.e., the cumulative regret, $R_{T}$). In other words, the speed of the learning is affected by the size of the decision space.

\subsection{Comparison of proposed algorithms}
To compare between the performance of the three proposed variants of the online learning solution, we perform an experiment considering DL model 1, DL model 2 and the available resources presented in Table \ref{tab:simulation_parameters}. We conduct the comparison in terms of the convergence of the average regret and how fast the probability of the optimal decision approaches a value of $1$. Due to the structure of the decision space resulting from the OLS algorithm, as the probability vector is updated at each time slot it will tend to be distributed among multiple optimal decisions rather than a single decision. On the other hand, since OLS-SA and OLS-RSA considers these decisions as sub actions of a single super action, the learning phase will result in a probability that is biased toward a single optimal super action. Therefore, for comparability of the three algorithms, at each time slot we add the probabilities corresponding to the multiple optimal decisions resulting from the OLS algorithm.

\begin{figure}
     \centering
     \begin{subfigure}[b]{0.23\textwidth}
         \centering
         \includegraphics[width=\textwidth]{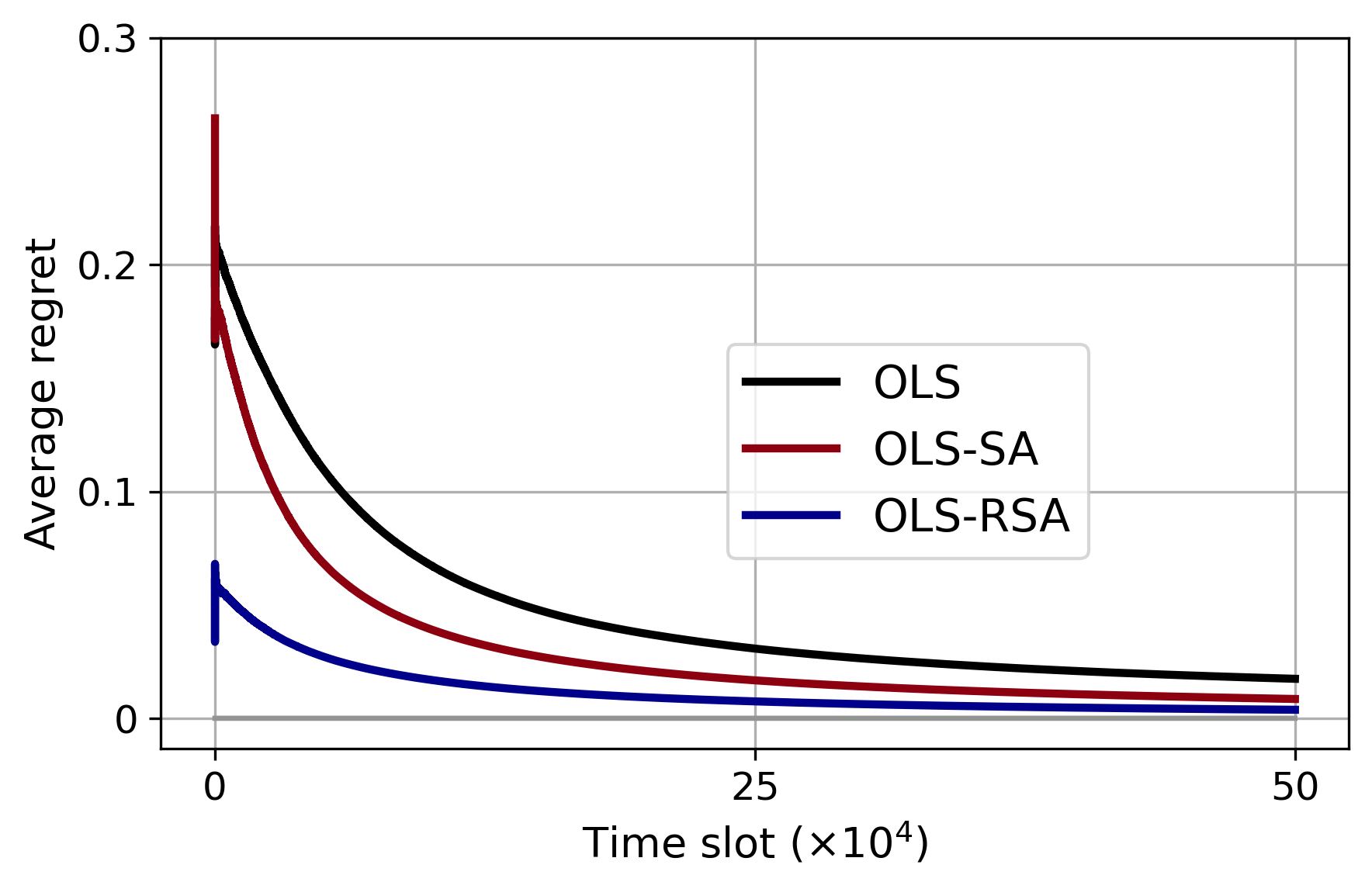}
         \caption{}
         \label{fig:regret_3_same}
     \end{subfigure}
     \begin{subfigure}[b]{0.23\textwidth}
         \centering
         \includegraphics[width=\textwidth]{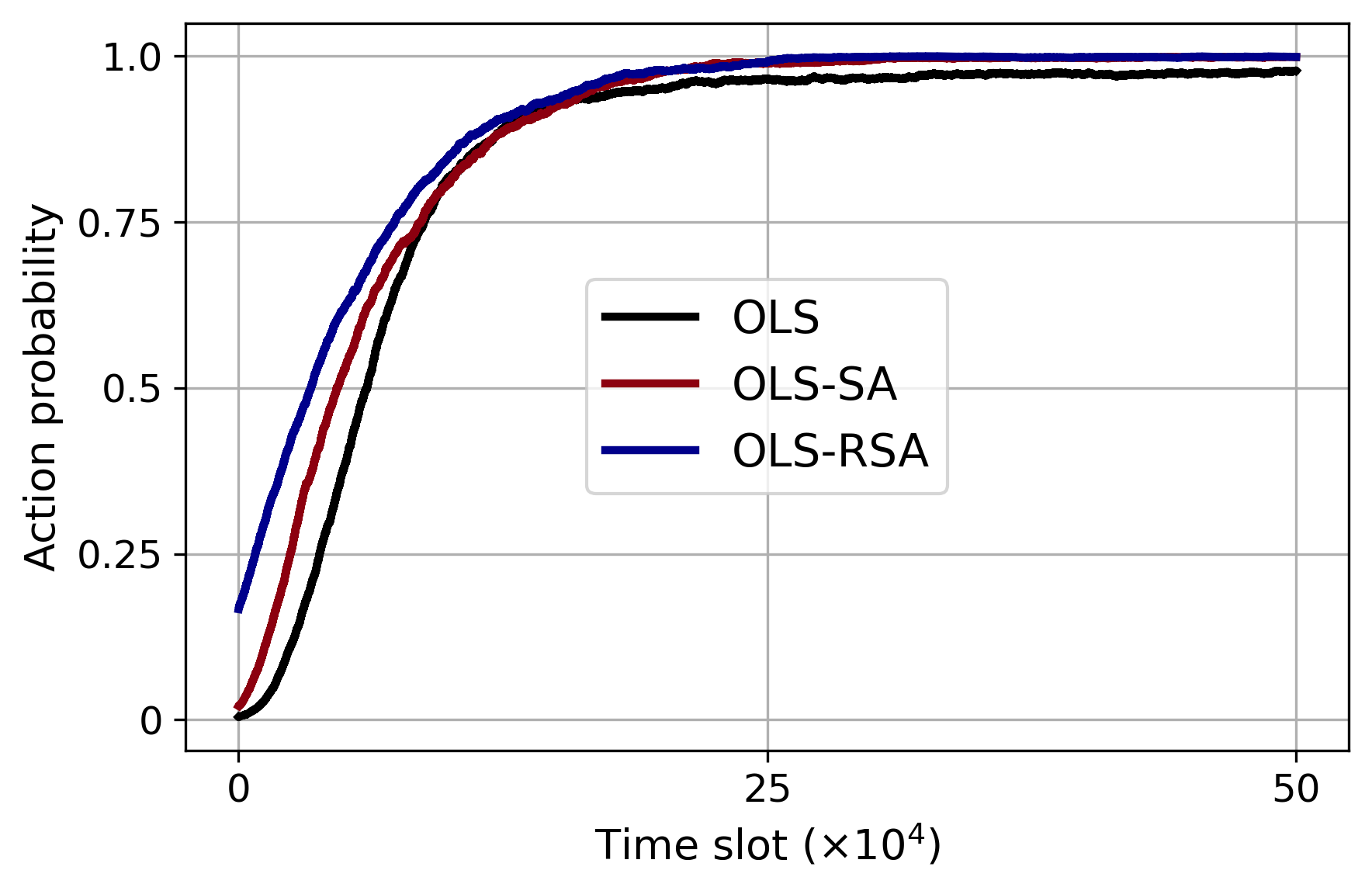}
         \caption{}
         \label{fig:weights_3_same}
     \end{subfigure}
     \begin{subfigure}[b]{0.23\textwidth}
         \centering
         \includegraphics[width=\textwidth]{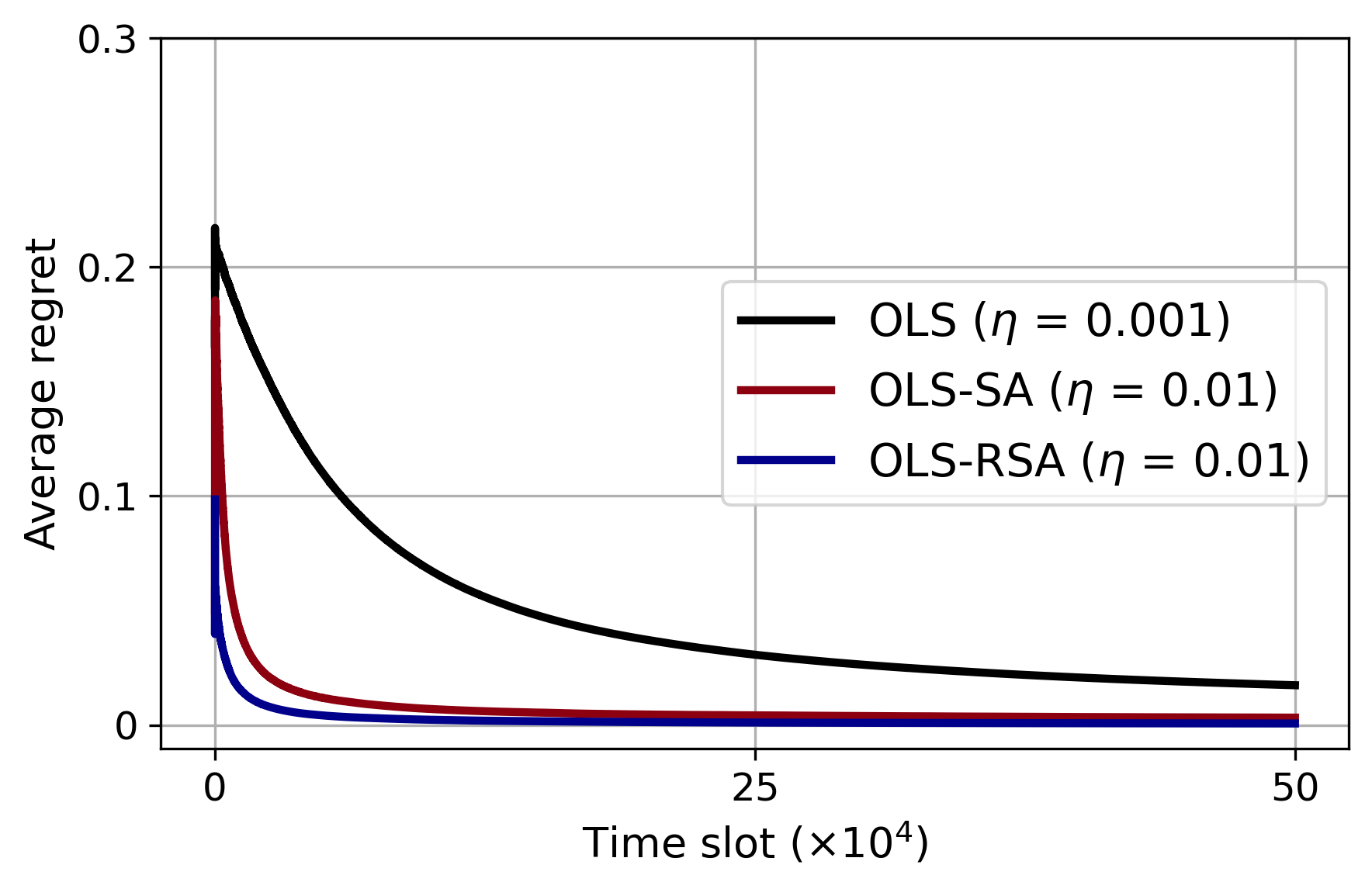}
         \caption{}
         \label{fig:regret_3_diff}
     \end{subfigure}
          \begin{subfigure}[b]{0.23\textwidth}
         \centering
         \includegraphics[width=\textwidth]{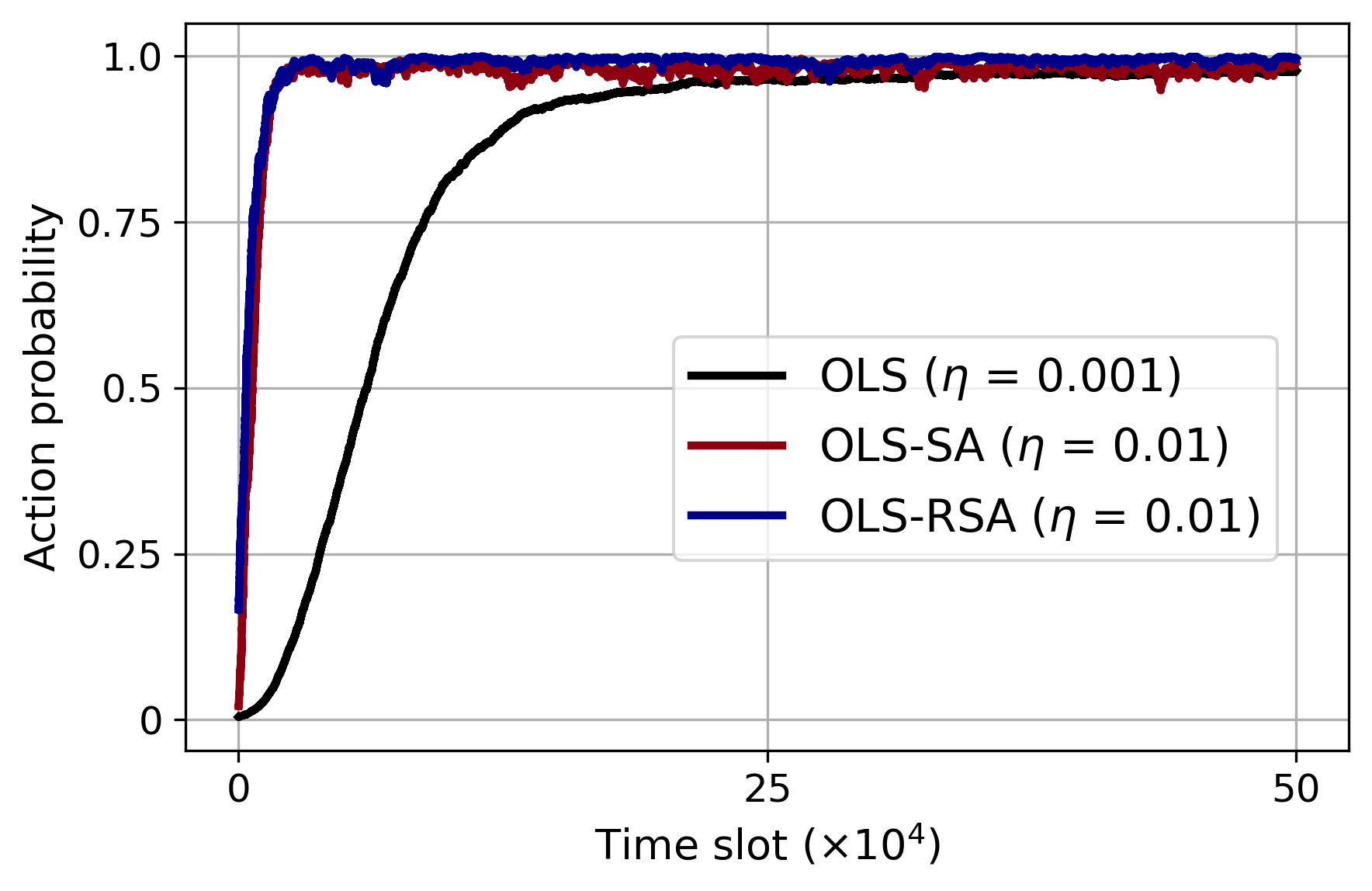}
         \caption{}
         \label{fig:weights_3_diff}
     \end{subfigure}
     \caption{Comparison between the three elimination methods: a) Average regret curve with $\eta = 0.001$, b) Action probability curve showing the probability given to the optimal action with $\eta = 0.001$ , c) Average regret curve using the best observable $\eta$, c) Action probability curve showing the probability given to the optimal action using the best observable $\eta$. }
     \label{fig:elimination_comparison}
\end{figure}
In Figure \ref{fig:elimination_comparison}, it is observed that in general the OLS-SA algorithm exhibits a lower cumulative regret and hence a faster convergence to a low average regret compared to the OLS algorithm. This is because by combining similar actions together into a super action, they undergo the same probability update, instead of being treated as independent. OLS-SA algorithm reduces the size of the decision space compared to OLS, which increases the speed of the learning. It is also observed that OLS-RSA exhibits the lowest cumulative regret of the three methods. This is because by reducing the size of the decision space, the online optimization controller is able to learn the decisions that maximizes the system performance faster. In addition to this, it is likely that the candidates of the optimal super action will result in a higher system performance compared to other actions, and hence resulting in a faster convergence to a low average regret. It is worth noting that both the OLS-SA and OLS-RSA algorithms exhibit a similar behavior in terms of the probability update of the optimal action (see Figure \ref{fig:elimination_comparison}-\subref{fig:weights_3_same}, and Figure \ref{fig:elimination_comparison}-\subref{fig:weights_3_diff}). For both algorithms, the probability given to the optimal action nearly converges to a value of $1$ as opposed to OLS. We highlight that OLS-SA and OLS-RSA result in a significant reduction in the action space compared to OLS. Therefore, we can achieve a remarkably faster convergence of the probability corresponding to the optimal super action to a value of $1$ at the best observable $\eta$ (see Figure \ref{fig:elimination_comparison}-\subref{fig:regret_3_diff}, and Figure \ref{fig:elimination_comparison}-\subref{fig:weights_3_diff}). 

\subsection{Time complexity analysis of the proposed algorithms}
Given the previous experimental setup, we compare between the cumulative time complexity of the three algorithms. To account for the time complexity of the pre-learning phase, we experimentally compute the number of times of executing the major operations described in Sec. \ref{sec:OL}.
\begin{figure}[t!]
	\centering
		\scalebox{1.4}{\includegraphics[width=0.33 \textwidth]{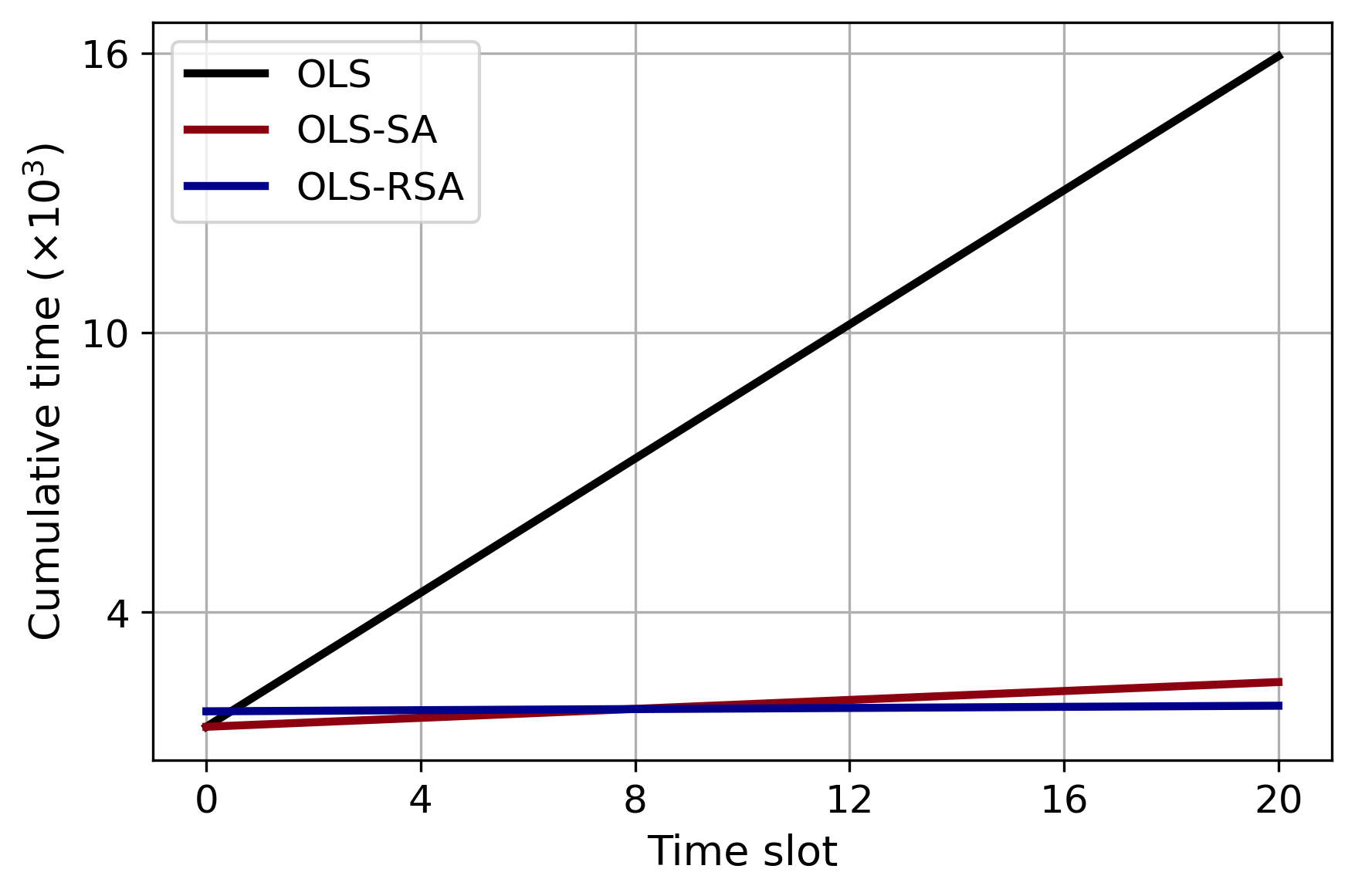}}
	\caption{Comparison of the cumulative time complexity given: $|\mathcal{A}_\text{OLS}| = 720$, $|\mathcal{A}_\text{OLS-SA}| = 48$, $|\mathcal{A}_\text{OLS-RSA}| = 6$, and $T = 20$  }
	\label{fig:complexity_analysis}
\end{figure}
In Figure \ref{fig:complexity_analysis}, we observe that OLS-RSA algorithm results in the highest initial overhead due to the steps of optimal super action candidacy checking. We also observe that the initial overhead of OLS and OLS-SA is the same. However, the OLS-SA and OLS-RSA algorithms reduce the size of the action space in the pre-learning phase compared to OLS. Hence, as time progresses the OLS algorithm results in a significant increase of cumulative time complexity. On the other hand, the OLS-RSA algorithm will start to exhibit the lowest cumulative time complexity as time progresses, since it results in a considerable decrease in the size of the decision space.

\subsection{Service performance evaluation}
We use the OLS-RSA algorithm to evaluate the individual performance of several DL models during the process of the online resource allocation. 
We conduct an experiment considering the four DL models in Table \ref{tab:services} and the simulation parameters in Table \ref{tab:simulation_parameters} with  $\eta = 0.001$. We set $l_\text{min$_{i}$} = 20$ $\forall i \in \mathcal{I}$, $m_\text{min$_{i}$} = 3$ $\forall i \in \mathcal{I}$, $C_\text{max$_{2}$} = 0.46$, $C_\text{max$_{3}$} = 0.38$, $D_\text{max$_{1}$} = 3.07$, $D_\text{max$_{2}$} = 3.07$, $D_\text{max$_{3}$} = 4.4$, $\Psi_\text{max} = 7$, $\Lambda_\text{max} = 7$. The action space is discretized according to the following:
\begin{itemize}
    \item $l_{{i}} \in \{20,55,80,100\}$ $\forall i \in \mathcal{I}$
    \item $m_{i} \in \{3,5,8,10\}$ $\forall i \in \mathcal{I}$
    \item $\psi_{i} \in \{1.5,1.8,2.2\}$ $\forall i \in \mathcal{I}$
    \item $\lambda_{i} \in \{1,2,3\}$ $\forall i \in \mathcal{I}$
\end{itemize}

\begin{figure*}
     \centering
     \begin{subfigure}[b]{0.325\textwidth}
         \centering
         \includegraphics[width=\textwidth]{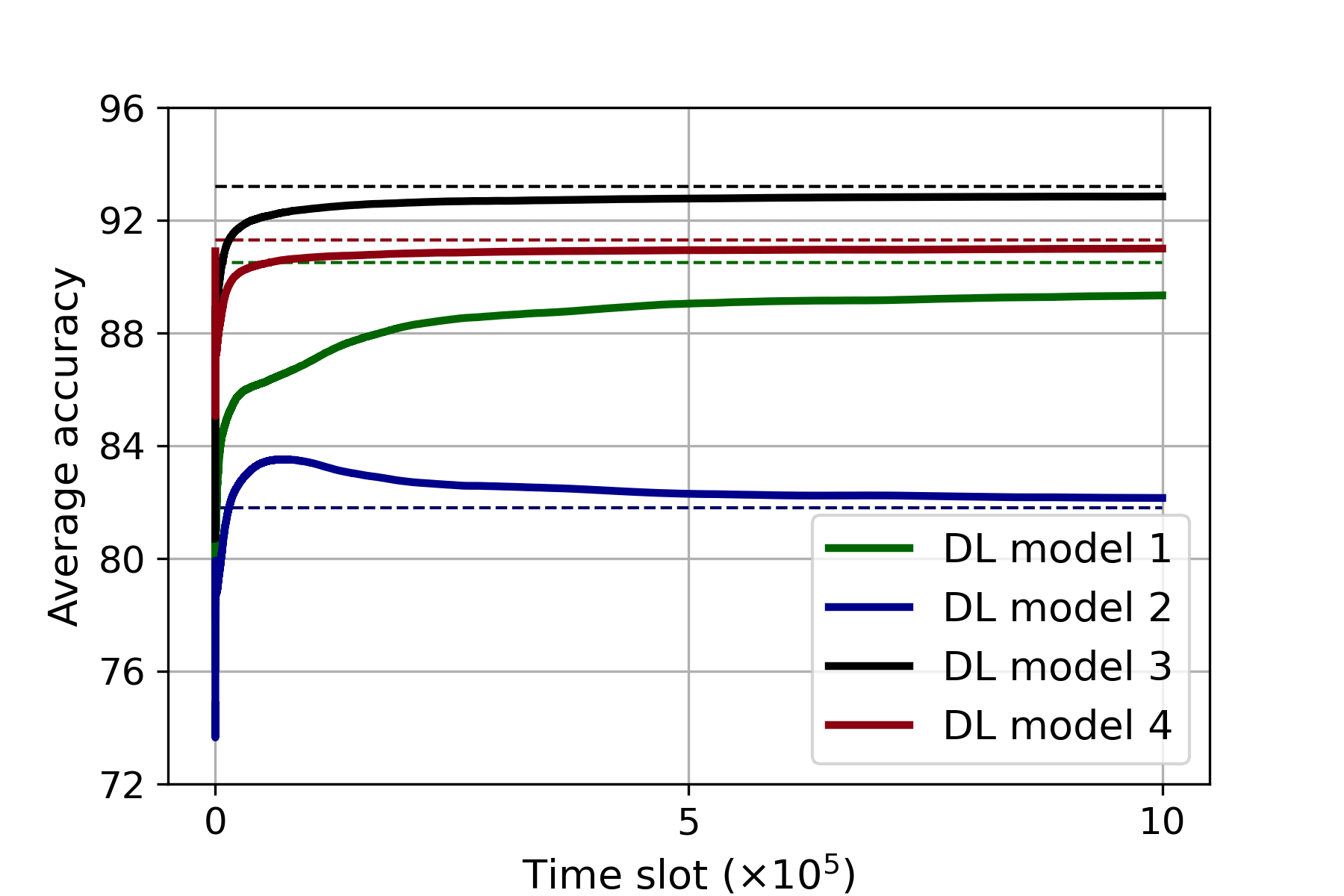}
         \caption{}
         \label{fig:service_perf}
     \end{subfigure}
     \begin{subfigure}[b]{0.325\textwidth}
         \centering
         \includegraphics[width=\textwidth]{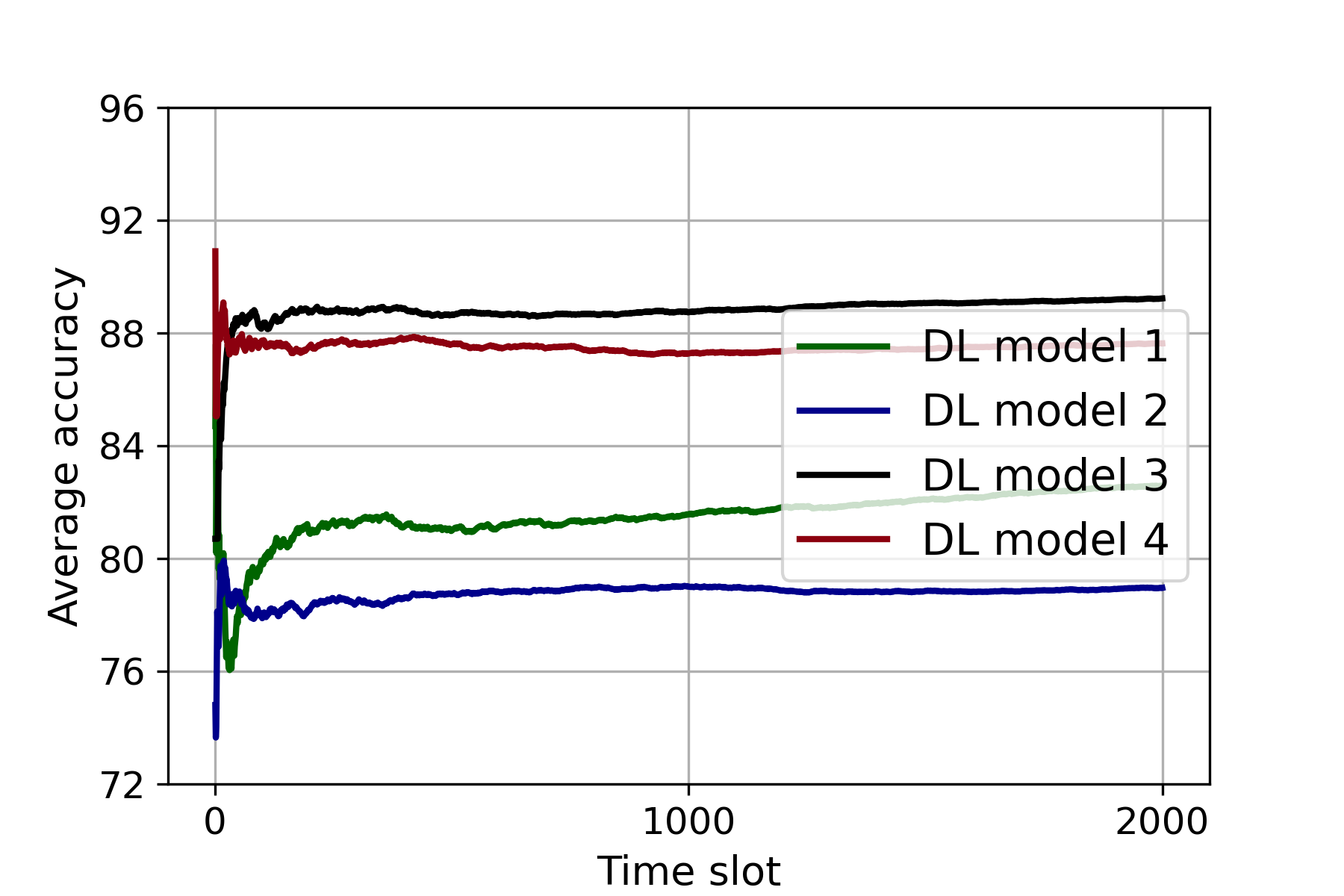}
         \caption{}
         \label{fig:service_perf_zoomed}
     \end{subfigure}
     \begin{subfigure}[b]{0.325\textwidth}
         \centering
         \includegraphics[width=\textwidth]{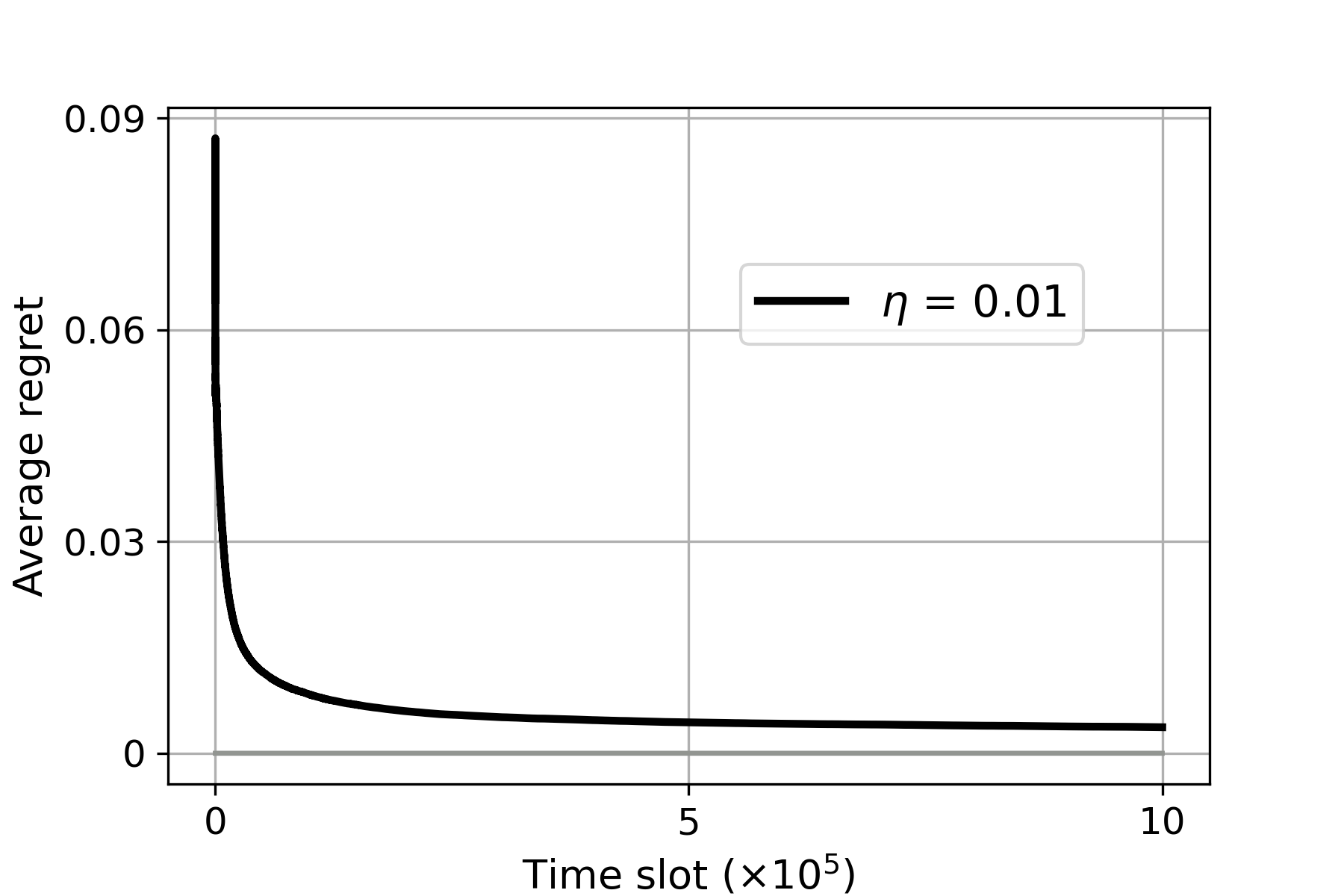}
         \caption{}
         \label{fig:service_perf_regret}
     \end{subfigure} 
     \caption{a) Average accuracy of each DL model, b) Experienced fluctuations during the initial time steps of the learning phase, c)Average regret curve}
     \label{fig:performance}
\end{figure*}

Analyzing the average accuracy of the DL models (see Figure \ref{fig:performance}), it is observed that each DL model experiences initial fluctuations until the average inference accuracy of each of them converges to a value which corresponds to the maximum of their summation. We observe a trade-off between the allocation of resources and hence the behavior of DL model 1 and DL model 2. This trade-off indicates that the allocation of resources cannot accommodate the combination of data size and number of epochs which leads to the highest possible inference accuracy for both DL models at the same time. It appears that since the objective of the problem is to maximize the summation of accuracies, this trade-off arises as more resources will be allocated to the DL model that weighs higher in terms of accuracy. Hence during the learning phase, the DL model which exhibits a lower accuracy may experience an initial increase in the average performance followed by a decrease as the online optimization controller learns the allocation which corresponds to the collective maximum performance.

\subsection{Biased subset selection evaluation}
\begin{figure}
     \centering
     \begin{subfigure}[b]{0.45\textwidth}
         \centering
         \includegraphics[width=\textwidth]{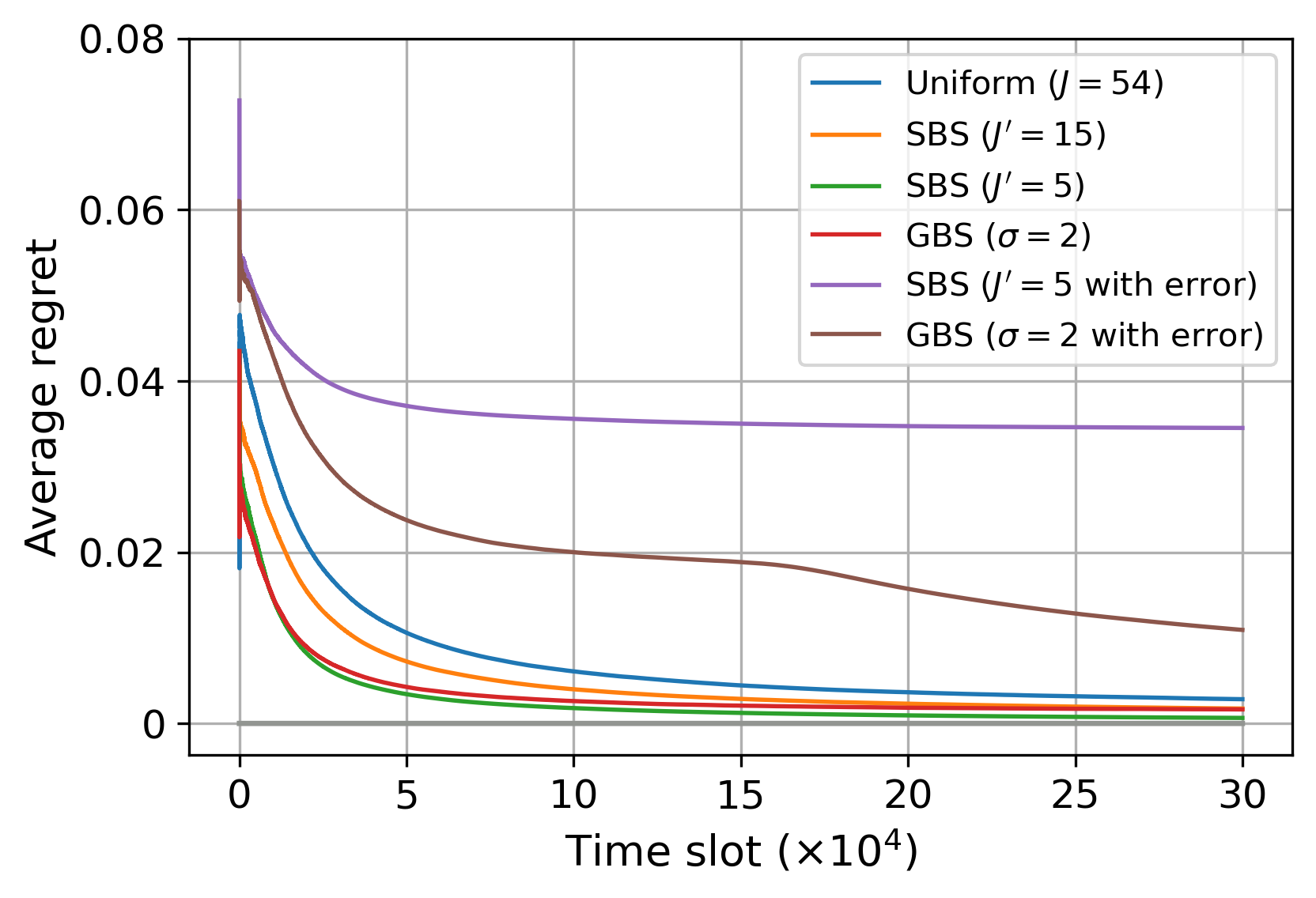}
         \caption{}
         \label{fig:Biased_schemes_regret}
     \end{subfigure}
     \begin{subfigure}[b]{0.45\textwidth}
         \centering
         \includegraphics[width=\textwidth]{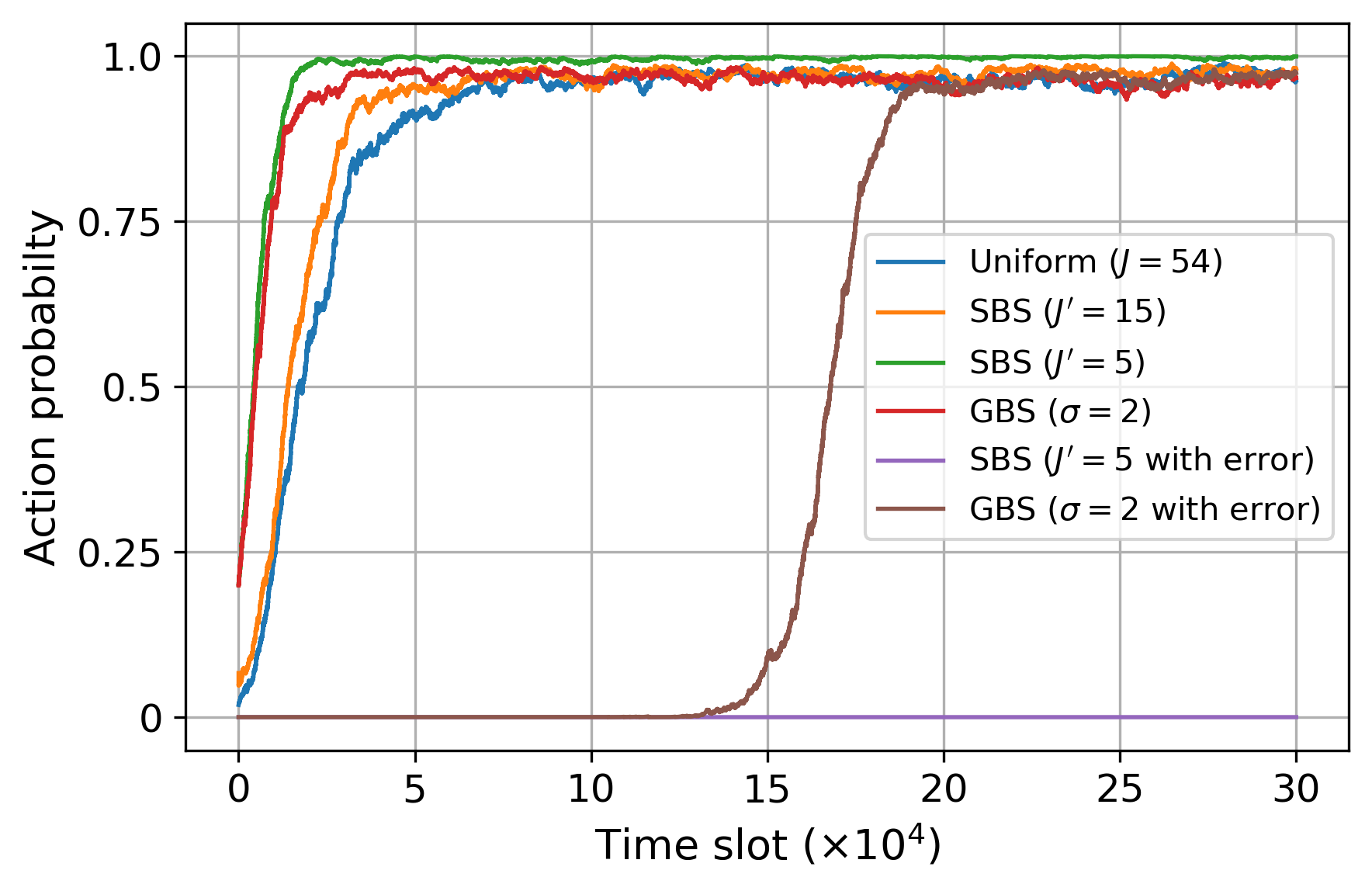}
         \caption{}
         \label{fig:Biased_schemes_prob}
     \end{subfigure} 
     \caption{Comparing the impact of different biased subset selection and initial probability distributions in terms of: a) The average regret b) The probability corresponding to the optimal action at each time step}
     \label{fig:Biased_distribution}
\end{figure}
We have been evaluating the performance considering an initial uniform probability distribution over the entire decision space. Considering the previous experimental setup with four DL models, we extend the analysis to compare between the impact of learning while considering a biased subset of the original decision space (i.e., SBS and GBS) as discussed in section \ref{sec:biased}, and learning with a uniform distribution over the entire decision space. We analyze the impact of the size of the considered subset $J'$ on the learning. We also study the behavior of the different biased subset selection approaches when it is identified with error and without error. We have identified that the optimal action corresponds to the index $j = 40$ in the decision space. Thus, we consider a strictly biased subset centred at $j = 40$, with $J' = 15$ and $J' = 5$ in the case without error, and another centred at $j = 9$ with $J' = 5$ for the case with error. We also consider a Gaussian biased subset in the case without error to be centred around the optimal action with $\mu = 40$, and another in the case with error with $\mu = 9$. 

In Figure \ref{fig:Biased_distribution}, it is observed that compared to a uniform distribution over the entire decision space, a biased subset selection which includes the optimal action results in a lower cumulative regret (i.e., smaller area under the average regret curve), and a faster convergence of the probability corresponding to the optimal action to a value of $1$. It is also observed that the smaller the size of the subset $J'$, the faster the convergence to the optimal action and hence a lower cumulative regret. In the case of identifying the biased subset of actions with error, we observe that the GBS approach allows the online optimization controller to learn the optimal super action in the long time. Accordingly, the average regret will start to decrease as the online optimization controller learns the optimal super action. Meanwhile, using the SBS approach, the learner is not able to identify the optimal super action since the probability assigned to it at initialization is equal to zero. In this case, the online optimization controller will learn a sub-optimal super action within the specified subset. Therefore, it is evident that the GBS approach performs better in comparison to SBS since it is able to account for error.


\section{Conclusions\label{sec:conclusion}}
Towards a new paradigm of 6G network slicing to support AI-based services, \textit{slicing for AI}, we proposed an online learning approach to jointly allocate resources and tune hyper-parameters for training AI models, while maximizing their collective performance. We formulated the problem as an accuracy maximization problem subject to resources, budget, and AI model training latency constraints. We presented an online learning solution with decision space reduction methods and biased decision space subset selection approaches.
The simulation results demonstrate the diverse capabilities and trade-offs of the implemented algorithms in achieving an optimal allocation decision while providing sub-linear regret guarantees. Moreover, our findings indicate that optimizing the action space leads to favorable outcomes in terms of faster convergence to the optimal solution, lower loss accumulation, and overall optimization of cumulative time complexity. Lastly, our results highlight the trade-offs in the proposed approaches of biasing a selected subset of the decision space.

\balance 

\bibliographystyle{IEEEtrannames}
\bibliography{SlicingForAI}

\end{document}